\documentstyle[aps,floats,epsf]{revtex}
\input amssym.def
\input amssym.tex
\begin{document}
\draft
\title{Towards a theory of the integer quantum Hall transition:
  continuum limit of the Chalker-Coddington model}
\author{Martin R. Zirnbauer}
\address{Institut f\"ur Theoretische Physik, Universit\"at zu K\"oln,
  Germany} \date{October 31, 1996} \maketitle
\begin{abstract}
  An $N$-channel generalization of the network model of Chalker and
  Coddington is considered.  The model for $N = 1$ is known to
  describe the critical behavior at the plateau transition in systems
  exhibiting the integer quantum Hall effect.  Using a recently
  discovered equality of integrals, the network model is transformed
  into a lattice field theory defined over Efetov's $\sigma$ model
  space with unitary symmetry.  The transformation is exact for all
  $N$, no saddle-point approximation is made, and no massive modes
  have to be eliminated.  The naive continuum limit of the lattice
  theory is shown to be a supersymmetric version of Pruisken's
  nonlinear $\sigma$ model with couplings $\sigma_{xx} = \sigma_{xy} =
  N/2$ at the symmetric point.  It follows that the model for $N = 2$,
  which describes a spin degenerate Landau level and the random flux
  problem, is noncritical.  On the basis of symmetry considerations
  and inspection of the Hamiltonian limit, a modified network model is
  formulated, which still lies in the quantum Hall universality class.
  The prospects for deformation to a Yang-Baxter integrable vertex
  model are briefly discussed.
\end{abstract}
\bigskip

\section{Introduction}

The single-electron states of a two-dimensional disordered electron
gas in a strong magnetic field are localized except at the energies of
the Landau band centers.  As the Fermi energy approaches such a band
center, a critical phenomenon takes place: the localization length
diverges and the Hall conductance jumps from one plateau to the next.
This phase transition, which belongs to the general class of Anderson
metal-insulator transitions, has been seen in several experiments and
a substantial amount of data on its critical behavior is available
from a number of numerical simulations (in the absence of
electron-electron interactions), see \cite{huckestein} and references
therein.  Unfortunately, in spite of considerable efforts expended
over the last decade, our analytical understanding of the
plateau-to-plateau transition is still rather poor.  It is expected
that the critical behavior is described by some nonunitary conformal
field theory, but this theory has not yet been identified.

There exist two opposite limits from which the transition in the
noninteracting system can be approached theoretically.  The first
limit is that of a slowly varying random potential with a correlation
length $l_c$ much larger than the magnetic length $l_B$.  In this
limit, the electron's motion can be described in semiclassical
terms \cite{trugman}.  More precisely, the motion separates into a
rapid cyclotron motion superimposed on a slow guiding center drift
along spatially localized equipotential lines.  As the Fermi energy
approaches the center of a Landau band, a percolating path develops
and a localization-delocalization transition takes place.  Close to
the transition, the quantum mechanical possibility for an electron to
tunnel from an equipotential to a neighboring one is a relevant
perturbation.  The essential features of this quantum percolation
transition were cast into a random network model by Chalker and
Coddington \cite{cc}.  In their model an electron acquires random ${\rm
  U}(1)$ phases while moving along the directed bonds of a square
network, and is scattered to the right or left every time it passes
through a node of the network.  The model will be reviewed in more
detail below.  Suffice it to say here that the model has been
studied by numerical simulation but has, in its original, spatially
isotropic formulation, defied analytical treatment up to now, although
a certain amount of analytical insight has come from consideration of
its anisotropic \cite{DHLee} and weak disorder \cite{ho_chalker}
versions.

The opposite limit is $l_c \ll l_B$.  Historically, this was the {\it
  first} limit to be studied analytically, the reason being that it is
this limit that was amenable to the field theoretic machinery
developed at the beginning of the 1980's by Wegner \cite{wegner},
Efetov \cite{efetov} and others.  Starting from a Gaussian white-noise
potential $(l_c = 0)$ Pruisken \cite{pruisken} used the replica trick
to set up a generating functional for the disorder averaged retarded
and advanced Green's functions of the single-electron Hamiltonian.
After a Hubbard-Stratonovitch transformation to matrix-valued fields,
he made a saddle-point approximation, valid in the limit of a high
Landau level.  This was followed by a gradient expansion leading to a
${\rm U}(n_+ + n_-) / {\rm U}(n_+)\times{\rm U}(n_-)$ nonlinear
$\sigma$ model with a vanishing number of replicas, $n_+ = n_- = 0$,
and a parity-violating topological term due to the strong magnetic
field.  The coupling constants of the model, $\sigma_{xx}$ and
$\sigma_{xy}$, were identified \cite{pruisken1} with the physical
conductivities of the 2d disordered electron gas.  The topological
coupling $\theta = 2\pi\sigma_{xy}$, by its very nature, has no effect
on the equations of motion of the classical field theory, but {\it
  does} change both the Hamiltonian and the symplectic structure and,
consequently, the commutator of the quantum theory.  It was
argued \cite{llp} that the nonlinear $\sigma$ model, while generically
being massive (i.e. having a finite correlation length) in two
dimensions, becomes critical at $\theta = \pi$.  The vanishing of the
mass gap corresponds to the appearance of delocalized states at the
center of a Landau band.  Thus, Pruisken's model provided the right
kind of scenario in which to develop a scaling theory of the
plateau-to-plateau transition.  A supersymmetric version of the model
first appeared in \cite{haw}.

Undeniably, Pruisken's nonlinear $\sigma$ model has served as a great
inspiration to theory.  One of its early successes was the
$\sigma_{xx}$-$\sigma_{xy}$ flow that was conjectured from
it \cite{Khmelnitskii,llp} and later verified by numerical and real
experiments, see \cite{huckestein} and references therein.  In spite of
this, Pruisken's model or, rather, its promoters have been criticized,
see \cite{mrz_iqhe} for a summary.  For one thing, the model has never
yielded any quantitative results for the critical behavior at the
transition, and much less has it been solved.  (The same statement
applies to the general class of nonlinear $\sigma$ models with a
topological term.  None of these has ever been solved, at least not
directly.)  For another, even the derivation of the model is
vulnerable to criticism: the validity of the saddle-point
approximation that is made to eliminate the so-called massive modes,
requires $\sigma_{xx} \gg 1$.  Although this inequality is satisfied
for the bare (or SCBA) value of $\sigma_{xx}$ in the limit of a high
Landau level, the renormalized theory at $\sigma_{xy} = 1/2$ is
expected to have a $\sigma_{xx}$ of order unity or less.  The cure
proposed by Pruisken was to {\it assume} the renormalizability of his
model, and appeal to the RG flow to take the coupling constant
$\sigma_{xx}$ from large to small values.  However, such an assumption
needs to be justified and, in fact, is not acceptable by current field
theoretic knowledge, for Pruisken's model apparently lacks the
conformal structure that is required of a fixed point theory with a
continuous symmetry in two dimensions. (In other words, the model,
while definitely being critical at $\sigma_{xy} = 1/2$, does not
possess the conservation laws expected of an infrared stable fixed
point.)

Two advances will be made in the present paper.  The first is to
establish a very close connection between Pruisken's nonlinear
$\sigma$ model and the network model of Chalker and Coddington at
criticality.  We will show that the latter can be viewed as a lattice
discretization of the former or, conversely, taking the continuum
limit of the network model yields the $\sigma$ model.  A more detailed
outline is the following.  We start out by reviewing the
supersymmetric version of Pruisken's model and the Chalker-Coddington
network model in Secs.~\ref{sec:pruisken} and \ref{sec:cc_model}.
Then, in Sec.~\ref{sec:susy}, the network model is reformulated as a
lattice-regularized field theory defined over Efetov's supersymmetric
nonlinear $\sigma$ model space with unitary symmetry.  This
reformulation is exact.  In contrast with the conventional method due
to Wegner, Efetov and others, no saddle-point approximation is needed
to eliminate the massive modes.  Moreover,
Sec.~\ref{sec:GL22_symmetry} shows that not only is the supersymmetric
reformulation of the network model defined over the same field space,
but it also has the same global symmetries as Pruisken's model.  At
the critical point of the network model, where the correlation (or
localization) length diverges, the long wave length physics of the
supersymmetric lattice theory is expected to be described by a
continuum field theory.  The symmetries dictate that this continuum
theory be Pruisken's model at the critical coupling $\sigma_{xy} =
1/2$.  As is argued in Sec.~\ref{sec:sig_xx}, the other coupling
constant, $\sigma_{xx}$, equals $1/2$.  The numerical value of
$\sigma_{xy}$ is checked in Sec.~\ref{sec:sig_xy}, by evaluating the
lattice action on a smooth field configuration with nonzero
topological charge.  Sec.~\ref{sec:N_channels} extends these results
to an $N$-channel network model with random ${\rm U}(N)$ matrices on
links.

What we learn from all this is that, although historically Pruisken's
model was first obtained from the Gaussian white noise limit $l_c \ll
l_B$, in a low Landau level it is actually more closely related to the
opposite limit $l_c \gg l_B$, since it is the latter that provides the
microscopic basis for the network model.

The mathematical basis underlying the exact transformation from the
$N$-channel network model to the supersymmetric lattice field theory
is quite natural and simple, and is briefly sketched as follows.
(Readers who are not interested in mathematical structures may want to
skip this paragraph.)  For a pair of positive integers $n,N$ consider
the tensor product ${\Bbb C}^n \otimes {\Bbb C}^N$, on which the group
${\rm GL}(nN)$ acts by linear transformations.  Span the corresponding
Lie algebra ${\rm gl}(nN)$ by bilinears $\{ {\bar c}_A^i c_B^j \}_{A ,
  B = 1, ..., n}^{i,j=1,...,N}$ in fermionic creators ${\bar c}$ and
annihilators $c$, which act in a Fock space with vacuum $| 0 \rangle$.
There exist two natural subalgebras: ${\rm gl}(n)$, generated by
$\sum_i {\bar c}_A^i c_B^i$, and ${\rm gl}(N)$, generated by $\sum_A
{\bar c}_A^i c_A^j$.  Now put $n = n_- + n_+$ and fill the ``negative
energy states'' to form $|{\rm vac}\rangle = \prod_{i=1}^N
\prod_{A=1}^{n_-} {\bar c}_{A}^{i} |0\rangle$.  The particle-hole
coherent states \cite{perelomov} that are generated by the action of
${\rm GL}(n)$ on $|{\rm vac}\rangle$, are holomorphic sections of a
line bundle associated to the homogeneous space $G/H := {\rm GL}(n) /
{\rm GL}(n_+)\times{\rm GL}(n_-)$ by the Slater-determinant
representation of ${\rm GL}(n_-)$ on $|{\rm vac}\rangle$.  They are
parameterized by a complex $n_+ \times n_-$ matrix $Z$ with adjoint
$Z^\dagger$.  By combining the closure relation for particle-hole
coherent states with a few elementary properties of Fermi-coherent (or
Grassmann-coherent) states, one can prove the following equality of
integrals:
       \begin{eqnarray}
       &&\int_{{\rm U}(N)} dU 
       \prod_{a=1}^{n_+} {\rm Det}
       \left( 1 - e^{+i\varphi_{+a}} U \right)
       \prod_{b=1}^{n_-} {\rm Det}
       \left( 1 - e^{-i\varphi_{-b}} U^\dagger \right)
       \nonumber \\
       &=& \int d\mu(Z,Z^\dagger) \
       {\rm Det}^{-N}\left( 1 + Z^\dagger Z \right)
       {\rm Det}^{N}\left( 1 + Z^\dagger e^{+i\varphi_+} Z 
       e^{-i\varphi_-} \right) ,
       \nonumber
       \end{eqnarray}
where $\varphi_\pm = {\rm diag}( \varphi_{\pm 1},...,\varphi_{\pm
  n_\pm})$ are diagonal matrices with real entries, $dU$ is the Haar
measure of a ${\rm U}(N)$ subgroup of ${\rm GL}(N)$, and $d\mu(Z,
Z^\dagger)$ expresses the ${\rm U}(n)$-invariant measure of a compact
symmetric space ${\rm U}(n)/{\rm U}(n_+)\times{\rm U}(n_-)$ contained
in the complex homogeneous space $G/H$.  This integral identity forms
the mathematical basis of our formalism.  Its supersymmetric
extension \cite{mrz_circular} permits to carry out the disorder average
over the random ${\rm U}(N)$ matrices placed on the links of the
network, at the expense of introducing an integration over fields $Z$
taking values in a symmetric superspace.
       
Another issue addressed in this paper is the question whether the
supersymmetric formulation of the network model offers the possibility
for an exact analytical solution.  Sec.~\ref{sec:vertex} reveals that
the model resembles what is called a {\it vertex model} in statistical
physics, in the sense that the Boltzmann weight is a product of
factors, one for each node, or vertex.  The weight associated with a
single vertex is called the {\rm R-matrix}.  The symmetries of the
R-matrix are investigated in Sec.~\ref{sec:symmetries}.  It is
eventually found that it can be interpreted as a map ${\cal R} : V
\otimes V \to V \otimes V$, where $V$ is an irreducible lowest-weight
module for the Lie superalgebra ${\rm gl}(n,n)$, and $n = n_+ + n_- =
1 + 1$ for the case of one retarded and one advanced Green's function.
This looks interesting as one may hope that ${\cal R}$ can be deformed
to an R-matrix that solves the quantum Yang-Baxter equation underlying
the integrability of two-dimensional vertex models.  One would then
have the possibility of an analytical and exact computation of
critical properties.  Unfortunately, the specific choice of local
directions for the single-particle motion on the network, shown in
Fig.~2 below, turns out to be {\it incompatible} with the standard
schemes \cite{chari_pressley} for constructing solutions of the quantum
Yang-Baxter equation.  The reason is that the R-matrix of a
Yang-Baxter solvable model always transfers from one side of the
vertex to the other, whereas the Chalker-Coddington vertex maps the
horizontal degrees of freedom into the vertical ones, or vice versa.
Thus the Chalker-Coddington model in its original form does not fit
into the canonical framework of the theory of integrable systems, and
I am not aware of any method to make analytical progress with it.

However, this is not yet the end of the story.  Additional insight can
be gained by considering the anisotropic limit \cite{DHLee} of the
Chalker-Coddington model.  This limit and its relation to Pruisken's
model are reviewed in Sec.~\ref{sec:anisotropic}.  Based on it,
Sec.~\ref{sec:integrable} proposes a modified version of the isotropic
one-channel network model, which differs from the original one in two
respects.  First, the direction of the single-particle motion does not
alternate constantly between being horizontal and vertical.  Instead,
an electron may either pass straight through a node (with {\it no}
change of direction), or else be scattered either to the right or to
the left.  Analysing such a model by the mapping onto Pruisken's
nonlinear $\sigma$ model, we find that it is likely {\it not} to be in
the quantum Hall universality class, but in a massive (Haldane type)
phase.  Therefore, a second modification is proposed, which is to add
a second channel of propagation on {\it half} the links, say the
horizontal ones, see Fig.~5b below.  By the mapping onto Pruisken's
model, the doubly modified network model is expected to be critical in
a range of values of the parameter characterizing the scattering at
the nodes.  Moreover, by the changed transfer dynamics, the standard
schemes for constructing solutions of the quantum Yang-Baxter equation
are no longer ruled out.  I hope to elaborate on this theme in a
future publication.

\section{Supersymmetric Formulation of Pruisken's Model (Definitions)}
\label{sec:pruisken}

The original formulation \cite{pruisken} of Pruisken's nonlinear
$\sigma$ model relied on the use of the replica trick.  When applied
to phenomena that are nonperturbative in the disorder strength, the
replica trick is not mathematically sound but has been demonstrated to
lead to incorrect results, at least in some instances \cite{vz}.  (The
analytic continuation to a vanishing number of replicas is not unique
in general.)  Fortunately, we can avoid the replica trick by using an
alternative, supersymmetric formalism \cite{efetov}, which is on firm
mathematical ground.  The purpose of this section is to briefly review
the supersymmetric formulation \cite{haw} of Pruisken's model in a
language that is well suited for what will follow below.  For
simplicity, only the model pertaining to one retarded and one advanced
Green's function is treated.  A more detailed discussion of the model
can be found in \cite{mrz_iqhe}.

To define Pruisken's nonlinear $\sigma$ model we first specify 
its field space, as follows.  Consider a pair $Z, {\tilde Z}$ of 
$2 \times 2$ complex supermatrices
        \[
        Z = \pmatrix{Z_{\rm BB} &Z_{\rm BF}\cr Z_{\rm FB} &Z_{\rm FF}\cr},
        \quad 
        {\tilde Z} = \pmatrix{{\tilde Z}_{\rm BB} &{\tilde Z}_{\rm BF}\cr 
        {\tilde Z}_{\rm FB} &{\tilde Z}_{\rm FF}\cr} ,
        \]
where the subscripts ${\rm B}$ and ${\rm F}$ stand for Bosonic and
Fermionic, and let $g = \pmatrix{A &B\cr C &D\cr} \in {\rm GL}(2|2)$
act on these by
        \begin{eqnarray}
        Z &&\mapsto g\cdot Z = (AZ+B)(CZ+D)^{-1} , 
        \nonumber \\
        {\tilde Z} &&\mapsto g\cdot {\tilde Z} = 
        (C+D{\tilde Z})(A+B{\tilde Z})^{-1} .
        \nonumber
        \end{eqnarray}
Following \cite{mrz_iqhe}, one identifies the pair $Z,{\tilde Z}$ as a
set of coordinates for the complex coset space $G / H := {\rm
  Gl}(2|2)/{\rm GL}(1|1)\times{\rm GL}(1|1)$, where the denominator is
the subgroup generated by the block-diagonal matrices $h = \pmatrix{A
  &0\cr 0 &D\cr}$.  With this identification, the action $Z \mapsto g
\cdot Z$ and ${\tilde Z} \mapsto g \cdot {\tilde Z}$
coincides \cite{mrz_iqhe} with the natural action of $G$ on $G / H$ by
left translation.  The supermatrices $Z$ and $\tilde Z$ then are
viewed as left-translates of the origin in $G/H$, by writing $Z \equiv
g\cdot 0 = BD^{-1}$ and $\tilde Z \equiv g\cdot {\tilde 0} = CA^{-1}$.

The coset space $G / H$, being a homogeneous space, admits only one
(up to multiplication by a constant) rank-two supersymmetric tensor
that is invariant under the action of $G = {\rm GL}(2|2)$.  In the 
coordinates $Z,\tilde Z$ it is given by
        \[
        \hat g = {\rm STr}(1-Z\tilde Z)^{-1}
        {\rm d}Z (1-\tilde ZZ)^{-1}{\rm d}\tilde Z ,
        \]
where ${\rm STr}$ denotes the supertrace.  The $G$-invariant
superintegration measure that derives from $\hat g$ is denoted
$D(Z,\tilde Z)$ and has the local\footnote{Superintegration measures,
also called integral superforms or Berezin forms, generically suffer
from a coordinate ambiguity, or anomaly, see \cite{rothstein}.  A
general procedure by which to define $D(Z,\tilde Z)$ {\it globally}
is described in Sec.~II~A of \cite{mrz_suprev}.} coordinate 
expression \cite{haw_mrz}
        \[
        D(Z,\tilde Z) = {\rm d}Z_{\rm BB} \wedge
        {\rm d}{\tilde Z}_{\rm BB} \wedge {\rm d}Z_{\rm FF} \wedge
        {\rm d}{\tilde Z}_{\rm FF} \ 
        {\partial^4 \over 
        \partial Z_{\rm BF} \partial {\tilde Z}_{\rm BF}
        \partial Z_{\rm FB} \partial {\tilde Z}_{\rm FB} } \ .
        \]
The integration domain for the bosonic variables is fixed by the conditions
        \[
        {\tilde Z}_{\rm FF} = - {\bar Z}_{\rm FF} , \quad
        {\tilde Z}_{\rm BB} = + {\bar Z}_{\rm BB} , \quad
        {\rm and} \quad
        | Z_{\rm BB} |^2 < 1 ,
        \]
where the bar denotes complex conjugation.  These conditions select a 
submanifold $M_{\rm B} \times M_{\rm F}$ of $G/H$, 
        \[
        M_{\rm B} = {\rm U}(1,1) / {\rm U}(1) \times {\rm U}(1) \simeq 
        {\rm H}^2 , \qquad
        M_{\rm F} = {\rm U}(2) / {\rm U}(1) \times {\rm U}(1) \simeq 
        {\rm S}^2 ,
        \]
on which the metric $\hat g$ is Riemann.  The variables $Z_{\rm FF}$
and $Z_{\rm BB}$ can be shown \cite{mrz_iqhe} to have a meaning as
complex stereographic coordinates on the two-sphere ${\rm S}^2$ and
two-hyperboloid ${\rm H}^2$, respectively.  The triple $(G/H,M_{\rm
  B}\times M_{\rm F}, \hat g)$ is a Riemannian symmetric
superspace \cite{mrz_suprev}, and is called Efetov's $\sigma$ model
space with unitary symmetry.  In most of the existing literature, this
space is parameterized by a $4\times 4$ supermatrix $Q$, introduced below.

After these preparations, Pruisken's nonlinear $\sigma$ model (or, rather, 
the supersymmetric version thereof) is defined by the functional integral
        \[
        \langle \bullet \rangle = \int {\cal D}(Z,{\tilde Z}) 
        \ \bullet \ \exp - S_{\rm cont}[Z,\tilde Z] \ ,
        \]
where $S_{\rm cont}[Z,\tilde Z] = \int L d^2r$ is obtained by integrating
the following Lagrangian:
        \begin{eqnarray}
        L &=& \sigma_{xx}^{(0)} ( {\cal L}_{xx} + {\cal L}_{yy} ) 
        + \sigma_{xy}^{(0)} ({\cal L}_{xy} - {\cal L}_{yx}) ,
        \nonumber \\
        {\cal L}_{\mu\nu} &=& {\rm STr}(1-Z\tilde Z)^{-1} \partial_\mu Z
        (1-\tilde ZZ)^{-1} \partial_\nu \tilde Z ,
        \label{lagrangian}
        \end{eqnarray}
and the functional integration measure is
        \[
        {\cal D}(Z,\tilde Z) = \prod_{{\bf r}=(x,y)}
        D\left( Z({\bf r}),{\tilde Z}({\bf r}) \right) .
        \]
By construction, this field theory is invariant under global ${\rm
  Gl}(2|2)$ transformations $Z({\bf r})\mapsto g\cdot Z({\bf r})$ and
${\tilde Z}({\bf r}) \mapsto g\cdot {\tilde Z}({\bf r})$ (each of
the terms ${\cal L}_{\mu\nu}$ is).  The partition function ${\cal Z}
= \langle 1 \rangle$ equals unity by supersymmetry.  To obtain physical 
observables, one adds sources and takes functional derivatives, as usual.

For some purposes it is useful to switch to a coordinate-independent 
language by introducing a $4\times 4$ supermatrix field $Q$ by 
        \[
        Q = \pmatrix{1 &Z\cr \tilde Z &1\cr} \pmatrix{1 &0\cr 0 &-1\cr}
        \pmatrix{1 &Z\cr \tilde Z &1\cr}^{-1} ,
        \]
in terms of which the Lagrangian takes the familiar form \cite{pruisken}
        \begin{eqnarray}
        {\cal L}_{xx} + {\cal L}_{yy} &=& {\rm STr}
        (\partial_x Q \partial_x Q + \partial_y Q \partial_y Q) / 8 =: L_0 ,
        \nonumber \\
        {\cal L}_{xy} - {\cal L}_{yx} &=& {\rm STr} ( \partial_x Q
        \partial_y Q - \partial_y Q \partial_x Q) Q / 8 =: L_{\rm top} .
        \nonumber
        \end{eqnarray}
The global action of ${\rm GL}(2|2)$ on $Q$ is $Q({\bf r}) \mapsto g
Q({\bf r}) g^{-1}$.  On a configuration manifold $C$ without boundary
the integral of $L_{\rm top}$ is quantized \cite{llp,haw_mrz}: $\int_C
L_{\rm top} d^2r = 2\pi i n$, and the integer $n$ is called the
``winding number'' or ``topological charge''.  $L_0$ and $L_{\rm top}$
are invariant under rotations in the $xy$ plane, and these are the 
only invariants that can be formed from ${\cal L}_{\mu\nu}$. Note that
$L_0$ is real-valued, whereas $L_{\rm top}$ is purely imaginary.
The coupling constants $\sigma_{xx}^{(0)}$ and $\sigma_{xy}^{(0)}$
have been interpreted \cite{pruisken1} as the longitudinal and Hall
conductivities of the 2d electron gas.

Let me mention in passing that the terms in the Lagrangian can be
written as
        \begin{eqnarray}
        L_0 &=& \left( {\partial^2 \over \partial x \partial x'} +
        {\partial^2 \over \partial y \partial y'} \right) \ln {\rm SDet}
        \left( 1 - Z(x,y)\tilde Z(x',y') \right)^{-1}\Big|_{x=x',
        y=y'} \ ,
        \nonumber \\    
        L_{\rm top} &=& \left( {\partial^2 \over \partial x \partial y'} -
        {\partial^2 \over \partial y \partial x'} \right) \ln {\rm SDet}
        \left( 1 - Z(x,y)\tilde Z(x',y') \right)^{-1}\Big|_{x=x',
        y=y'} \ ,
        \label{kaehler}
        \end{eqnarray}
where ${\rm SDet}$ denotes the superdeterminant.  The function
$\ln{\rm SDet}(1-Z\tilde Z)$ is called the K\"ahler potential of the
K\"ahler supermanifold $G / H$ parameterized by the complex
coordinates $Z$ and $\tilde Z$.

\section{The Chalker-Coddington Model}
\label{sec:cc_model}

The Chalker-Coddington model was formulated \cite{cc} for the purpose
of describing the plateau-to-plateau transition in systems exhibiting
the integer quantum Hall effect.  Its microscopic foundation as a 
model for the motion of a single 2d electron subject to a smooth random
potential and a strong magnetic field, is explained in the original
paper.

The model is defined on a finite or infinite square lattice forming a
network of nodes and directed links, see Fig.~1.  A ``wave function''
of the model is a collection of complex amplitudes $\{ \psi(l) \}$,
one for each link $l$ of the network.  The dynamics is governed not by
a Hamiltonian but by a unitary operator $U$, called the one-step time
evolution operator. $U$ acts on wave functions, and is given by a
sequence of two distinct operations.  The first one is stochastic and,
in a microscopic picture, accounts for the random phase acquired by the
electron's guiding center while drifting along equipotentials between
saddle points of the random potential.  In formal language, the
stochastic phases are encapsulated in a unitary operator $U_0$ that is
diagonal in the basis provided by the links: $U_0(l',l) =
\delta(l',l)^{\vphantom{+}} \exp i\varphi(l)$, where $\varphi(l)$ are
uncorrelated random variables with a uniform distribution on the
interval $[0,2\pi]$.
\begin{figure}
\hspace{2.5cm}
\epsfxsize=10cm 
\epsfbox{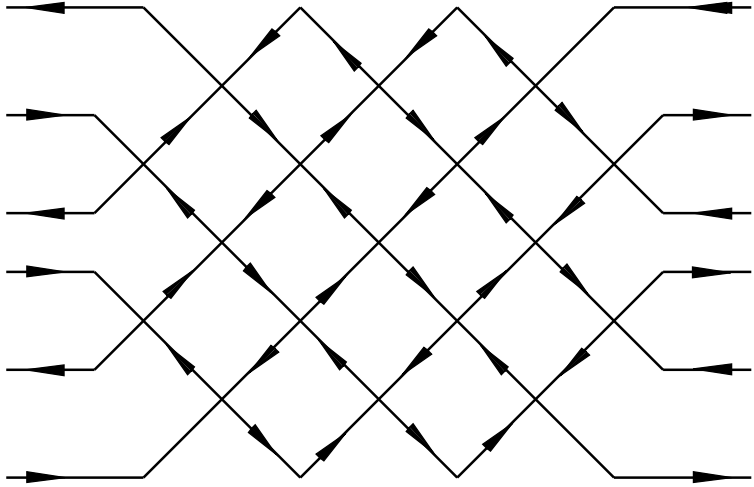}
\vspace{10pt}
\caption{The network model of Chalker and Coddington in a quasi 1d geometry.
A single electron propagates ballistically along the links and is scattered
at the nodes of the network.}
\end{figure}

The second process built into the one-step time evolution operator is
deterministic (i.e., nonrandom) and accounts for the quantum
mechanical possibility for an electron to tunnel across a saddle
point.  This is modeled via the nodes of the network.  One imagines
that the guiding center drift motion of the electron follows the
direction indicated by the arrows in Fig.~1.  An electron incident on
a node can be scattered either to the left or to the right,
corresponding to the two possibilities of continuing on its way along
the same equipotential, or tunneling to a neighboring one.  It cannot
be backscattered, and it cannot pass straight through a node.  The
probabilities for scattering to the right or left are denoted $p$ and
$1-p$.  The probability {\it amplitudes} for scattering at the nodes
then have magnitude $\sqrt{p}$ and $\sqrt{1-p}$ respectively.  They
are taken to be real, but they cannot all be chosen positive as this
would violate unitarity.  Various choices are possible.  For
definiteness we take the amplitudes to be negative for right-up to
left-up and for right-down to left-down turns (see Fig.~1) and
positive otherwise.  (Which choice is made actually turns out to be
immaterial.  All we need is that some choice consistent with unitarity
exists.) All this fixes a unitary operator $U_1$ with matrix elements
$U_1(l',l)$ that vanish unless the scattering process $l \to l'$ is
allowed by the (one-step) dynamics, in which case $U_1(l',l) = \pm
\sqrt{p}$ or $U_1(l',l) = \pm\sqrt{1-p}$, as specified.  The full
one-step time evolution operator is the product $U = U_1 U_0$.

For the purpose of doing numerical simulations, one usually takes the
network to be a long strip (quasi-1d geometry) with periodic boundary
conditions for the short direction to minimize finite size effects.
By computing the exponential growth of the transfer matrix for the
strip and averaging over many realizations of the disorder embodied by
$U_0$, one extracts a Lyapunov exonent or inverse localization length.
For $p = 1/2$ the localization length is found to grow linearly with
the width of the strip, indicating a critical point.

The existence of critical behavior at $p = 1/2$ can be anticipated by
the following argument \cite{cc}.  Consider the Chalker-Coddington
model with the open boundary conditions of Fig.~1.  For $p = 0$ (left
turns only) all electron states encircle elementary squares of the
network in the counterclockwise direction, and thus are strongly
localized.  The same statement applies for $p = 1$ (right turns only),
except that now the orientation of the motion around squares is
clockwise, and an {\it edge state} at the boundary of the strip
appears, as is seen by inspection of Fig.~1.  The appearance of an
edge state implies that somewhere in the interval $p \in [0,1]$ a
delocalized state must form.  For symmetry reasons, this is expected
to happen at the left-right symmetric point $p = 1/2$.

\begin{figure}
\hspace{2.5cm}
\epsfxsize=10cm
\epsfbox{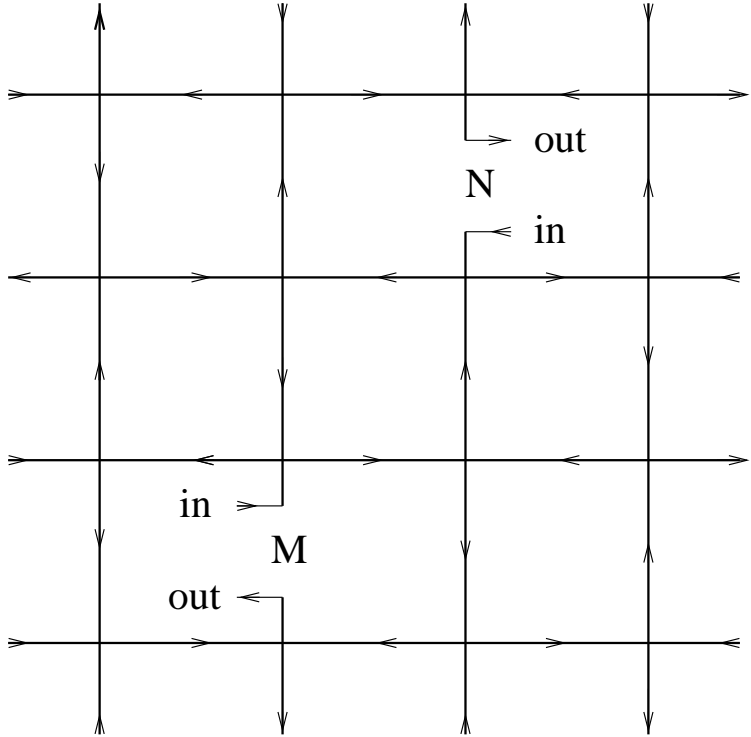}
\vspace{10pt}
\caption{Chalker-Coddington network with two interior contacts $M$ and
$N$.  The corresponding links $l_M$ and $l_N$ are connected to one 
incoming and one outgoing channel each.}
\end{figure}

The method to be introduced in Sec.~\ref{sec:susy} is quite capable of
dealing with a quasi-1d geometry and its boundary conditions.
However, for maximal simplicity we will consider below a somewhat
different setup, where periodic boundary conditions are imposed in
both directions, i.e. the network is placed on a torus ${\rm S}^1
\times {\rm S}^1$.  We envisage making a conductance measurement
between two {\it interior} contacts $M$ and $N$ as shown in Fig.~2.
These contacts are ``point'' contacts in the sense that they each
attach only to a {\it single} link, $l_M$ and $l_N$.  For most
purposes it is helpful to imagine that each link with a contact is
replaced by two links, an ``in-link'' feeding the network from a
reservoir, and an ``out-link'' draining the network through an
outgoing channel leading back to the reservoir.  Outgoing-wave
boundary conditions are imposed on out-links, i.e.  probability flux
that is scattered into such a link by the action of $U_1$, must exit
and never returns to the network.  The conductance pertaining to two
interior contacts $M$ and $N$ can be computed from the
Landauer-B\"uttiker formula, $g_{MN} = |S_{MN}|^2$, where $S_{MN}$ is
the S-matrix element relating an incoming state on the in-link of
$l_N$ to an outgoing wave amplitude on the out-link of $l_M$.  This
S-matrix element is determined by the solution of Schr\"odinger's
equation $U\psi = e^{i\alpha}\psi$ with incoming-wave boundary
conditions at $N$.  The eigenphase $e^{i\alpha}$ may be gauged away by
a global shift of the random ${\rm U}(1)$ factors on links, $U_0(l)
\mapsto e^{i\alpha} U_0(l)$.  It is convenient to implement the loss
of probability flux through out-links by the modification $U_0(l) =
\exp i\varphi(l) \to U_0(l^{\rm out}) \equiv 0$ for every link $l$
with a contact.  Then, solving the equation $\psi = U \psi$ by
iteration, we easily see that the d.c. conductance equals
        \[
        g_{MN} = |S_{MN}|^2 = \Big| \langle l_M^{\rm out} | T | 
        l_N^{\rm in} \rangle \Big|^2 ,
        \]
where $T$ is the operator
        \begin{equation}
        T = U_1 + U_1 U_0 U_1 + U_1 U_0 U_1 U_0 U_1 + ...
        = \left( 1 - U_1 U_0 \right)^{-1} U_1 .
        \nonumber
        \end{equation}
Note that owing to $U_0(l_M^{\rm out}) = U_0(l_N^{\rm out}) = 0$, the
eigenvalues of $U_1 U_0$ lie {\it inside} the unit circle (rather than
on the unit circle) in ${\Bbb C}$, so the inverse $(1-U_1 U_0)^{-1}$
is well-defined.  Without changing the matrix element $|S_{MN}|^2$, we
set $U_0(l_M^{\rm in}) = U_0(l_N^{\rm in}) = 0$.

\section{Reformulation by Supersymmetry}
\label{sec:susy}

In this section we will show how to map the problem of calculating the
disorder averaged conductance $\langle g_{MN} \rangle$ on an
equivalent lattice field theory.  {}From $g_{MN} = \Big| \langle
l_M^{\rm out} | T | l_N^{\rm in} \rangle \Big|^2$, $T = (1-U_1
U_0)^{-1} U_1$, and $T^\dagger = (1-U_1^\dagger U_0^\dagger)^{-1}
U_1^\dagger$, the mean conductance $\langle g_{MN} \rangle$ equals the
disorder average of the expression
        \[ 
        \sum_{l_m l_n} \langle l_N^{\rm in} 
        | (1 - U_1^\dagger U_0^\dagger)^{-1} | l_m \rangle \ 
        U_1^\dagger (l_m , l_M^{\rm out}) \ \langle l_M^{\rm out} | 
        (1 - U_1 U_0)^{-1} | l_n \rangle \ U_1 (l_n , l_N^{\rm in}) .  
        \] 
This disorder average can be processed by a variant of Efetov's 
supersymmetry method introduced in \cite{mrz_circular}, as follows.  
The method starts by expressing the product of matrix elements as a 
Gaussian superintegral:
        \begin{eqnarray} 
        &&\langle l_N^{\rm in} | (1 - U_1^\dagger U_0^\dagger)^{-1} |
        l_m \rangle \langle l_M^{\rm out} | (1 - U_1 U_0)^{-1} | l_n \rangle 
        \nonumber \\ 
        &=& \int \prod_l D\left( \psi(l) , \bar\psi(l) \right) \
        \psi_{-{\rm B}}(l_N^{\rm in}) \bar\psi_{-{\rm B}}(l_m) 
        \psi_{+{\rm B}} (l_M^{\rm out}) \bar\psi_{+{\rm B}}(l_n) 
        \nonumber \\ 
        &&\hspace{2cm} \times \exp \Big( - \bar\psi_{+\sigma}(l') 
        \big[ \delta(l',l) - U_1(l',l) U_0(l) \big]
        \psi_{+\sigma}(l) 
        \nonumber \\ 
        &&\hspace{4cm} - \bar\psi_{-\tau}(l')
        \big[ \delta(l',l) - {\bar U}_1(l,l') {\bar U}_0(l) \big]
        \psi_{-\tau}(l) \Big) ,
        \nonumber 
        \end{eqnarray} 
where $\psi_{\pm\sigma}$ are the components of a complex superfield $l
\mapsto \psi(l)$.  The index $\sigma = {\rm B}$ or ${\rm F}$
distinguishes between Bosonic and Fermionic components, and $\pm$
relates to retarded ($T$) and advanced ($T^\dagger$) Green's
functions.  The bar denotes complex conjugation.  $D(\psi,\bar\psi)$
denotes the ``flat'' superintegration measure, i.e., is given by the
product of the differentials of the bosonic components times the
product of partial derivatives with respect to the fermionic ones. The
summation convention is used here and throughout the paper.  

For the following step, it is notationally convenient to absorb $U_1$
temporarily by setting $\hat\psi_{+\sigma}(l) = \bar\psi_{ +
  \sigma}(l')U_1(l',l)$ and $\hat\psi_{-\tau}(l) = \bar\psi_{-\tau}
(l') {\bar U}_1(l,l')$.  We will now trade the average over random
phases $\int \prod_l dU_0(l) = \int \prod_l {\rm d}\varphi(l) / 2\pi$
for an integral over Efetov's $\sigma$ model space with unitary
symmetry, reviewed in Sec.~\ref{sec:pruisken}.  This is done by a kind
of Hubbard-Stratonovitch transformation,
        \begin{eqnarray}
        &&\int {\prod_l}' dU_0(l) \exp \left( \hat\psi_{+\sigma}(l) U_0(l)
        \psi_{+\sigma}(l) + \hat\psi_{-\tau}(l) {\bar U}_0(l) \psi_{-\tau}(l)
        \right)
        \nonumber \\
        &=& \int {\prod_l}' D\mu \left( Z(l) , {\tilde Z}(l) \right) \exp
        \left( \hat\psi_{+\sigma}(l) Z_{\sigma\tau}(l) \psi_{-\tau}(l) + 
        \hat\psi_{-\tau}(l) {\tilde Z}_{\tau\sigma}(l) \psi_{+\sigma}(l) 
        \right) ,
        \label{HS}
        \end{eqnarray}
which is a special case of a more general result proved in
\cite{mrz_circular}.  The $2\times 2$ supermatrix valued fields $l
\mapsto Z(l)$ and $l \mapsto \tilde Z(l)$ are lattice discretizations
of the continuous fields of Pruisken's model.  The superintegration
measure,
        \[
        D\mu(Z,\tilde Z) := {\rm const} \times
        D(Z,\tilde Z) \ {\rm SDet}(1-\tilde Z Z) ,
        \]
is normalized by the condition $\int D\mu(Z,\tilde Z) = 1$.  Because
of the boundary condition at the contacts, $U_0(l_M) = U_0(l_N) = 0$
(in- and out-links), the product over links on both sides of
(\ref{HS}) excludes $l_M$ and $l_N$.  An alternative scheme of
implementing the boundary condition is to let the product run over
{\it all} links, including the boundary ones, and set
        \begin{equation}
        Z(l_M) = \tilde Z(l_M) = Z(l_N) = \tilde Z(l_N) = 0 .
        \label{contacts}
        \end{equation}
We adopt this scheme.  By using (\ref{HS}) to deal with the disorder
average of the product of matrix elements, we obtain
        \begin{eqnarray}
        &&\Big\langle \langle l_N^{\rm in} | 
        (1 - U_1^\dagger U_0^\dagger)^{-1}
        | l_m \rangle \langle l_M^{\rm out} 
        | (1 - U_1 U_0)^{-1} | l_n \rangle \Big\rangle 
        \nonumber \\
        &=& \int \prod_l D\mu\left( Z(l) , \tilde Z(l) \right)
        \int \prod_l D\left( \psi(l) , \bar\psi(l) \right) \ 
        \psi_{-{\rm B}}(l_N^{\rm in}) \bar\psi_{-{\rm B}}(l_m) \psi_{+{\rm B}}
        (l_M^{\rm out}) \bar\psi_{+{\rm B}}(l_n) 
        \nonumber \\
        &&\hspace{1.5cm} \times \exp \Big( - \bar\psi_{+\sigma}(l) 
        \psi_{+\sigma}(l) + \bar\psi_{+\sigma}(l') U_1(l',l) 
        Z_{\sigma\tau}(l) \psi_{-\tau}(l) 
        \nonumber \cr
        &&\hspace{3cm} - \bar\psi_{-\tau}(l) \psi_{-\tau}(l)
        + \bar\psi_{-\tau}(l') {\bar U}_1(l,l') {\tilde Z}_{\tau\sigma}(l) 
        \psi_{+\sigma}(l) \Big) .
        \nonumber
        \end{eqnarray}
The final step is to carry out the Gaussian integral over $\psi$ and
$\bar\psi$, which gives
        \begin{eqnarray}
        \langle g_{MN} \rangle &=& \int \prod_l D\mu \left( Z(l) , 
        {\tilde Z}(l) \right) \ {\rm SDet}_{\cal H} \left( 1 - Z_U 
        \tilde Z \right)^{-1} 
        \nonumber \\
        &&\times \Big[ Z_U(1-\tilde Z Z_U)^{-1}\Big]_{\rm BB}
        (l_M^{\rm out},l_M^{\rm out}) \times \Big[ {\tilde Z}_{U^{\dagger}}
        (1 - Z {\tilde Z}_{U^{\dagger}})^{-1}
        \Big]_{\rm BB} (l_N^{\rm in},l_N^{\rm in}) ,
        \nonumber
        \end{eqnarray}
by an elementary calculation.  Here $Z_U$ denotes the supermatrix
field $Z$ evolved forward in time by one action of the deterministic
scattering operator $U_1$, i.e. $Z_U = U_1^{\vphantom{\dagger}} Z 
U_1^{\dagger}$ or, in more explicit notation,
        \[
        \left(U_1^{\vphantom{\dagger}}Z U_1^\dagger\right)_{\sigma\tau}(l,l') 
        = U_1(l,l'') Z_{\sigma\tau}(l'') {\bar U}_1(l',l'') .
        \]
Similarly, ${\tilde Z}_{U^{\dagger}}$ is the supermatrix field
${\tilde Z}$ evolved backward in time by one step: ${\tilde
  Z}_{U^\dagger} = U_1^{\dagger} {\tilde Z} U_1^{\vphantom{\dagger}}$.
The subscript ${\cal H}$ on ${\rm SDet}$ indicates that the
superdeterminant runs over both superspace and the Hilbert space of
wave functions supported on links.

In summary, we have shown that the average conductance pertaining to 
two interior contacts $M$ and $N$ can be computed as a two-point 
function $\langle g_{MN} \rangle = \langle {\cal O}(M) \tilde{\cal O}(N) 
\rangle$ of operators
        \begin{eqnarray}
        {\cal O}(M) &=& \Big[ Z_U ( 1 - {\tilde Z} Z_U )^{-1} \Big]
        (l_M^{\rm out} , l_M^{\rm out}) ,
        \nonumber \\
        \tilde{\cal O}(N) &=& \Big[ {\tilde Z}_{U^{\dagger}} ( 1 - Z 
        {\tilde Z}_{U^{\dagger}} )^{-1} \Big]
        (l_N^{\rm in} , l_N^{\rm in}) ,
        \nonumber
        \end{eqnarray}
in a lattice field theory
        \[
        \langle \bullet \rangle = \int \prod_l D \left( Z(l) , {\tilde Z}(l)
        \right) \ \bullet \ \exp - S_{\rm latt}[Z,\tilde Z] ,
        \]
with lattice action
        \begin{equation}
        S_{\rm latt} [ Z , \tilde Z ] = \ln {\rm SDet}_{\cal H}
        (1 - Z_U \tilde Z) - \ln {\rm SDet}_{\cal H}(1-Z\tilde Z) .
        \label{lattice_action}
        \end{equation}
The second term in $S_{\rm latt}$ originates from the measure
$D\mu(Z,\tilde Z) = D(Z,\tilde Z) \ {\rm SDet}(1-Z{\tilde Z})$. 
Note that the functional integral for the two-point function 
$\langle {\cal O}(M) \tilde{\cal O}(N) \rangle$ is regularized
by the boundary condition (\ref{contacts}). 

What we have done is an {\it exact rewriting} of the original problem.
It might superficially look as though we have made the problem more
complicated by transforming from ${\rm U}(1)$ phase integrals to an
integral over supermatrix fields $Z$ and $\tilde Z$, but this is not
so.  The point is that while the ${\rm U}(1)$ phases $U_0(l)$ {\it
  fluctuate independently}, the dominant contributions to the integral
over the supermatrix field come from {\it slowly varying} field
configurations.  The latter property is best seen by combining terms
to rewrite the lattice action as follows:
        \[
        S_{\rm latt}[Z,\tilde Z] = \ln {\rm SDet}_{\cal H} 
        \left( 1 - (1-\tilde Z Z)^{-1} {\tilde Z} ( Z_U - Z) \right) .
        \]
The right-hand side vanishes for constant fields ($Z_U = Z$) and is
small for slowly varying ones.  This will allow us to take a continuum
limit and describe the low energy or long wave length physics of the
Chalker-Coddington network model by a continuum field theory.  In view
of (\ref{kaehler}) and (\ref{lattice_action}), it should not come as a
surprise that this field theory will turn out to be Efetov's nonlinear 
$\sigma$ model with unitary symmetry, augmented by Pruisken's topological 
term.

The formula we have derived for $\langle g_{MN} \rangle$ illustrates
how the transport coefficients of the network model can be expressed
as correlations functions of a supersymmetric lattice field theory.
In the remainder of the paper we will not discuss the specific
correlator $\langle g_{MN} \rangle$ any further, but will concentrate
on the general structure of the theory.

\section{Global symmetry under ${\rm GL}(2|2)$}
\label{sec:GL22_symmetry}

To get ready for taking the continuum limit, we will now elucidate the
symmetries of the lattice field theory.  We are going to show that 
it is invariant under global transformations
        \begin{eqnarray}
        Z(l) &&\mapsto g\cdot Z(l) = \big( AZ(l)+B \big)
        \big( CZ(l)+D \big)^{-1} , 
        \nonumber \\
        {\tilde Z}(l) &&\mapsto g\cdot {\tilde Z}(l) = 
        \big( C+D\tilde Z(l) \big) \big( A+B\tilde Z(l) \big)^{-1} ,
        \nonumber
        \end{eqnarray}
for $g = \pmatrix{A &B\cr C &D\cr} \in {\rm GL}(2|2)$.  The integration
measure $D\left( Z(l) , \tilde Z(l) \right)$ has this invariance 
{\it by definition}.  To see that the lattice action (\ref{lattice_action})
is invariant, we first transform the factor ${\rm SDet}\left( 1 - 
Z(l) \tilde Z(l) \right)$ for a single link $l$, temporarily dropping
the argument $l$ for notational simplicity:
        \begin{eqnarray}
        &&{\rm SDet}\left( 1 - (g\cdot Z)(g\cdot \tilde Z)\right)
        \nonumber \\
        &=& {\rm SDet} \left( 1 - (AZ+B)(CZ+D)^{-1}(C+D\tilde Z)
        (A+B\tilde Z)^{-1}\right)
        \nonumber \\
        &=& {\rm SDet} \left( 1 - (Z+A^{-1}B)(1+D^{-1}CZ)^{-1}
        (\tilde Z+D^{-1}C)(1+A^{-1}B\tilde Z)^{-1}\right) .
        \nonumber
        \end{eqnarray}
This expression is further processed by setting $A^{-1}B =: X$ and
$D^{-1}C =: \tilde X$ and using the identity
        \begin{eqnarray}
        &&1 - (Z+X)(1+\tilde X Z)^{-1}(\tilde Z + \tilde X)(1+X\tilde Z)^{-1}
        \nonumber \\
        &=& (1-X\tilde X)(1+Z\tilde X)^{-1}(1-Z\tilde Z)(1+X\tilde Z)^{-1} ,
        \nonumber
        \end{eqnarray}
which follows from elementary algebra.  We then get
        \[
        {\rm SDet}\left( 1 - (g\cdot Z)(g\cdot\tilde Z)\right) = 
        {\rm SDet}\left( (1-X\tilde X)(1+Z\tilde X)^{-1}(1-Z\tilde Z)
        (1+X\tilde Z)^{-1}\right) .
        \]

The ``dynamical'' factor ${\rm SDet}_{\cal H}(1-Z_U \tilde Z)^{-1}$ is 
transformed in an identical fashion.  Because the transformation
is global and $U_1$ acts as the identity in superspace, we have
        \[
        (g\cdot Z)_U^{\vphantom{+}} = U_1^{\vphantom{-1}} (g\cdot Z) U_1^{-1}
        = g \cdot (U_1^{\vphantom{-1}} Z U_1^{-1}) = g \cdot Z_U ,
        \]
i.e., the actions of $g$ and $U_1$ commute.  We thus obtain
        \begin{eqnarray}
        &&{\rm SDet}_{\cal H}\left( 1 - (g\cdot Z)_U^{\vphantom{+}} 
        (g\cdot Z)\right)^{-1}
        = {\rm SDet}_{\cal H}\left( 1 - (g\cdot Z_U)(g\cdot\tilde Z)
        \right)^{-1}
        \nonumber \\
        &=& {\rm SDet}_{\cal H}\left( (1-X\tilde X)(1+Z_U\tilde X)^{-1}
        (1-Z_U\tilde Z)(1+X\tilde Z)^{-1}\right) .
        \nonumber
        \end{eqnarray}
By combining factors and using the multiplicativity of the 
superdeterminant, we arrive at
        \[
        S_{\rm latt} [g\cdot Z,g\cdot\tilde Z] = 
        S_{\rm latt} [Z,\tilde Z] - \ln {\rm SDet}_{\cal H}(1+Z\tilde X)^{-1}
        + \ln {\rm SDet}_{\cal H}(1+Z_U\tilde X)^{-1} .
        \]
The last two terms on the right-hand side cancel each other by
        \[
        {\rm SDet}_{\cal H}(1 + Z_U\tilde X) = 
        {\rm SDet}_{\cal H}(1 + U_1^{\vphantom{-1}} Z U_1^{-1} \tilde X)
        = {\rm SDet}_{\cal H}(1+Z\tilde X) ,
        \]
since $\tilde X$ is constant on links and therefore commutes with
$U_1$ (i.e. $U_1^{-1} \tilde X U_1^{\vphantom{-1}} = \tilde X$).  This
establishes the global invariance $S_{\rm latt}[g\cdot Z,g\cdot\tilde
Z] = S_{\rm latt}[Z,\tilde Z]$.

The global ${\rm GL}(2|2)$ symmetry of $S_{\rm latt}$ is broken by the
boundary condition (\ref{contacts}), of course.  Such a symmetry
breaking is needed for the mathematical consistency of the
formulation, since the integral over the $q = 0$, or zero momentum,
mode of the noncompact bosonic sector would otherwise be divergent.

In Sec.~\ref{sec:susy} we saw that the supersymmetric reformulation of
the Chalker-Coddington model is defined over the same field space as
Pruisken's model.  What we have just learned is that these models
not only share the field space, but also have the {\it same internal
  symmetries}.  This is already a strong indication of their
equivalence at the critical point $p = 1/2$ resp.  $\sigma_{xy}^{(0)}
= 1/2$.

\section{Reformulation as a Vertex Model}
\label{sec:vertex}

Further analysis is facilitated by the observation that the
``Boltzmann weight'' of the lattice field theory,
        \[
        W[Z,\tilde Z] = {\rm SDet}_{\cal H} (1- \tilde Z Z_U)^{-1}
        {\rm SDet}_{\cal H}(1-\tilde Z Z) ,
        \]
factors to a large extent.  This factorization is seen as follows.
Consider the expression $\tilde Z(l'') U_1^{\vphantom{-1}}(l'',l')
Z(l') U_1^{-1}(l',l)$ and take $l$ to be the link $l = D$, say,
emanating in the downward direction from the node shown in Fig.~3a.
The matrix element $U_1^{-1}(l',D)$ is nonzero only if $l'$ is one of
the two links, $l' = L$ or $l' = R$ in the figure, that flow into the
same node.  In either case, the matrix element $U_1(l'',l')$ leads to
the final state $l''$ being one of only two links, namely $l'' = l =
D$ (the one we started from), or else $l'' = U$, the link emanating
from the node in the upward direction.  Similarly, if the initial
state is $l = U$, the final state is $l'' = U$ or $l'' = D$.  These
statements remain true (mutatis mutandi) if the node of Fig.~3a is
replaced by any other node of the network.  (Half of the nodes of the
network are obviously equivalent to the one considered.  The other
half of them has links emanating in the horizontal direction and nodes
flowing into it in the vertical direction, see Fig.~3b.  These nodes
become equivalent to the one of Fig.~3a after a rotation by ninety
degrees.)
\begin{figure}
\hspace{2.0cm}
\epsfxsize=13cm
\epsfbox{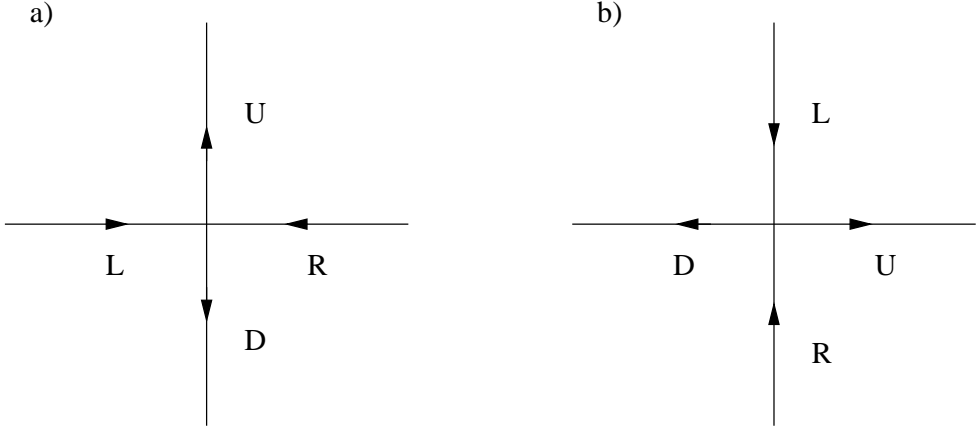}
\vspace{10pt}
\caption{The two types of elementary vertex of the Chalker-Coddington model.
Type b) is obtained from type a) by a rotation through $\pi/2$.}
\end{figure}

Thus, $\tilde Z U_1^{\vphantom{-1}} Z U_1^{-1}$ viewed as an operator
in the space of states supported on links, connects only {\it pairs of
links}, namely the two links that emanate from one and the same node.
This means that the Boltzmann weight $W$ is a product of factors,
one for each node:
        \begin{equation}
        W[Z,\tilde Z] = \prod_{{\rm nodes} \ n} {\cal R} 
        \left( \tilde Z(U_n) , \tilde Z(D_n) , Z(L_n) , Z(R_n) \right) ,
        \label{vertex}
        \end{equation}
where the assignment of link labels $U,D,L,R$ is defined by Fig.~3.
{}From the definition of $U_1$ in Sec.~\ref{sec:cc_model}, the weight 
for a single node works out to be
        \begin{eqnarray}
        &&{\cal R} \left( \tilde Z(U) , \tilde Z(D) , Z(L) , Z(R) \right) = 
        \nonumber \\
        &&{\rm SDet} \left( [1-\tilde Z(U)Z(U)][1-\tilde Z(D)Z(D)]
        [1-\tilde Z(L)Z(L)][1-\tilde Z(R)Z(R)] \right)^{1/2}
        \label{R_matrix} \\
        \times&&{\rm SDet}^{-1} \pmatrix{
        1-p\tilde Z(U)Z(R) - (1-p)\tilde Z(U)Z(L)
        &\sqrt{p(1-p)} \tilde Z(U)[Z(L)-Z(R)] \cr
        \sqrt{p(1-p)} \tilde Z(D)[Z(L)-Z(R)]
        &1-p\tilde Z(D)Z(L) - (1-p)\tilde Z(D)Z(R) \cr} .
        \nonumber
        \end{eqnarray}
A model of the kind (\ref{vertex}) is called a {\it vertex model} in
statistical mechanics, and ${\cal R}$ is called the R-matrix.  The
global symmetries of this R-matrix will be investigated in
Sec.~\ref{sec:symmetries}.  In Sec.~\ref{sec:integrable} we will 
show how to pass from the integration over the fields $Z,\tilde Z$
to a summation over superspin degrees of freedom. 

\section{Continuum Limit}
\label{sec:continuum}

For values of the parameter $p$ close to 1/2, where the
Chalker-Coddington model undergoes a quantum percolation transition,
the correlation (or localization) length is very large, and we expect
the lattice functional integral to be dominated by fields that vary
slowly.  Our goal in this section is to extract from
(\ref{lattice_action}) the continuum field theory governing these
slowly varying modes.

Recall the following facts: (i) Pruisken's model and the
supersymmetric (SUSY) reformulation of the Chalker-Coddington model are
defined over the very same complex field space, $G / H = {\rm GL}(2|2)
/ {\rm GL}(1|1) \times {\rm GL}(1|1)$.  (ii) Both the action
functional of Pruisken's model, $S_{\rm cont}$, and the lattice action
of the SUSY reformulated Chalker-Coddington model, $S_{\rm latt}$, are
invariant under global ${\rm GL}(2|2)$ transformations.  (iii) The
metric tensor of the field space $G / H$,
        \[
        {\hat g} = {\rm STr}(1-Z\tilde Z)^{-1} {\rm d}Z
        (1-\tilde ZZ)^{-1}{\rm d}\tilde Z = - {\rm STr}({\rm d}Q)^2 / 8,
        \]
is invariant under ${\rm GL}(2|2)$ [i.e. under $Q \mapsto g Q g^{-1}$
with $g \in {\rm GL}(2|2)$], and this is the {\it only} second-rank
supersymmetric tensor on the homogeneous space $G / H$ that has this
invariance.  When combined with a standard field theoretic power
counting argument, these facts lead to the conclusion that the
continuum limit of the lattice action $S_{\rm latt}$ must be a linear
combination of the four terms ${\cal L}_{\mu\nu}$ $(\mu , \nu = x ,
y)$ in (\ref{lagrangian}), which are induced from $\hat g$.

A further constraint comes from the spatial symmetries.  At large
scales (i.e. in the infrared limit), where details of the network
structure are washed out, the Chalker-Coddington model acquires an
invariance under rotations in space.  {}From the four terms ${\cal
  L}_{\mu\nu}$ one can form only two linear combinations that are
rotationally invariant.  These are precisely the ones that figure in
the expression (\ref{lagrangian}) for Pruisken's Lagrangian (Pruisken
et al. \cite{llp} arrived at them by essentially the same
argument), namely ${\cal L}_{xx} + {\cal L}_{yy}$ and ${\cal L}_{xy} -
{\cal L}_{yx}$.  We thus conclude that the continuum limit of the SUSY
reformulated Chalker-Coddington model at criticality $(p = 1/2)$ is
Pruisken's model at the critical topological coupling $\theta = \pi =
2\pi\sigma_{xy}^{(0)}$, and some as yet unknown coupling
$\sigma_{xx}^{(0)}$:
        \[
        S_{\rm latt} [ {\rm smooth} \ {\rm fields} ] = S_{\rm cont} 
        \Big|_{ \sigma_{xx}^{(0)} = ? \ , \ \sigma_{xy}^{(0)} = 1/2} \ .
        \]
The only thing that remains to be done then, is to work out the
numerical value of the coupling constant $\sigma_{xx}^{(0)}$.  In the
following subsection we will argue that
        \[
        \sigma_{xx}^{(0)} = 1/2 \qquad ({\rm for} \ p = 1/2) .
        \]
The small value of $\sigma_{xx}^{(0)} = 1/2$ means that the theory is
at {\it strong coupling}.  (The weak-coupling limit, where Pruisken's
derivation is valid, is $\sigma_{xx}^{(0)} \gg 1$.)  The superscript
$(0)$ alerts us to the fact that this is a ``bare'' value, obtained by
a naive continuum limit, i.e. without taking into account possible
renormalizations coming from modes with short wave lengths.

In the following we pay no attention to the boundary conditions and
assume the network to be infinitely extended.

\subsection{The coupling constant $\sigma_{xx}^{(0)}$}
\label{sec:sig_xx}

To determine the value of $\sigma_{xx}^{(0)}$ for the SUSY
reformulated Chalker-Coddington model, we first expand $S_{\rm latt}$
to quadratic order in $Z$ and $\tilde Z$, and then take the long wave
length limit.  The quadratic part of $S_{\rm latt}$ is conveniently
obtained by expanding the expression (\ref{R_matrix}) for the
R-matrix:
        \begin{eqnarray}
        &&- 2 \ln {\cal R} 
        \left( {\tilde Z}(U) , {\tilde Z}(D) , Z(L) , Z(R) \right) 
        \nonumber \cr
        &&\hspace{0.5cm} = 
        {\rm STr} \Big( {\tilde Z}(U) \big[ Z(U) - 2(1-p) Z(L) \big]
        + \big[ {\tilde Z}(R) - 2p {\tilde Z}(U) \big] Z(R)
        \nonumber \cr
        &&\hspace{1cm} + {\tilde Z}(D) \big[ Z(D) - 2(1-p) Z(R) \big] + 
        \big[ {\tilde Z}(L) - 2p {\tilde Z}(D) \big] Z(L) \Big)
        + {\cal O}(Z^2 {\tilde Z}^2) .
        \nonumber
        \end{eqnarray}
Now recall the meaning of the labels $U,D,L,R$ defined by Fig.~3a, and
let $(n_x , n_y) \in {\Bbb Z}^2$ be the Cartesian coordinates of the
node in that figure.  Then, if $a$ is the lattice constant of the
network,
        \begin{eqnarray}
        Z(L) &=& Z(a n_x - a / 2 , a n_y) , \quad
        Z(D) = Z(a n_x , a n_y - a / 2 ) ,
        \nonumber \cr
        Z(R) &=& Z(a n_x + a / 2 , a n_y) , \quad
        Z(U) = Z(a n_x , a n_y + a / 2 ) .
        \nonumber
        \end{eqnarray}
Similar expressions hold for the other type of node, shown in Fig.~3b
(rotate by $\pi/2$).  Summing $\ln{\cal R}$ over all nodes of the
lattice we get an expression for the quadratic part $S_2$ of $S_{\rm
  latt}$, which is of the general form
        \[
        S_2 = \sum_{ll'} \left( \delta(l,l') - T(l,l') \right)
        {\rm STr} {\tilde Z}(l) Z(l) .
        \]
The matrix $T(l,l')$ represents a bistochastic process, i.e. it 
obeys the sum rules
        \[
        \sum_l T(l,l') = 1 = \sum_{l'} T(l,l') .
        \]
The first of these means that, if $\{P_t(l)\}$ is a set of dynamical
variables with dynamics $P_{t+1}(l) = \sum_{l'} T(l,l') P_t(l')$, then
the total probability is conserved: $\sum_l P_t(l) = {\rm const}$
(independent of time $t$).  It is not hard to guess the physical
meaning of this: the dynamics at hand is nothing but the {\it
  classical approximation} to the full quantum dynamics of the
Chalker-Coddington model.  In other words, expanding the lattice
action $S_{\rm latt}$ to second order in $Z, \tilde Z$ is equivalent
to neglecting all quantum interference effects and approximating the
Chalker-Coddington model by its classical limit.

The classical model has been looked at in a number of papers, most
recently in Sec.~I~B of \cite{xrs}, where it was shown that the
dynamics at large wave lengths simplifies to classical diffusion, with
the longitudinal conductivity of the network being $\sigma_{xx}^{(0)}
= \sqrt{p(1-p)}$.  (There is also a nonvanishing Hall conductivity
$\sigma_{xy}^{(0)} = p$.)  By translating this result into the present
context, we obtain for $S_2$ a diffusive action,
        \[
        \sqrt{p(1-p)} \int {\rm STr} \left( 
          - \tilde Z \partial_x^2 Z
          - \tilde Z \partial_y^2 Z \right) d^2r ,
        \]
for large wave lengths.  Matching this expression to the quadratic part
of $S_{\rm cont}$, Eq.~(\ref{lagrangian}), gives $\sigma_{xx}^{(0)} =
1/2$ at $p = 1/2$, as announced above.

\subsection{The coupling constant $\sigma_{xy}^{(0)}$}
\label{sec:sig_xy}

The critical points of the Chalker-Coddington model and Pruisken's
model are at $p = 1/2$ and $\sigma_{xy}^{(0)} = 1/2 \ ({\rm mod} \ 1)$
respectively.  Therefore, as was argued earlier, there cannot be any
doubt that the continuum limit of the lattice action
(\ref{lattice_action}) at $p = 1/2$ is Pruisken's action with
topological coupling $\sigma_{xy}^{(0)} = 1/2$.  Nevertheless, it is
both reassuring and instructive to check this by direct calculation,
which is what we do next.

In Sec.~\ref{sec:sig_xx} the value of $\sigma_{xx}^{(0)}$ was found by
looking at the quadratic part of the action.  The topological 
coupling $\sigma_{xy}^{(0)}$, by its very nature, evades any
such attempt at perturbative calculation.  To extract it from
(\ref{lattice_action}) a different, nonperturbative scheme must be
used.  A direct approach would be to perform a gradient expansion
around a topologically nontrivial background.  For reasons that will
be explained at the end of Sec.~\ref{sec:symmetries}, this is not easy
to do.  Here we will follow a different procedure, which is to
evaluate $S_{\rm latt}$ on the lattice discretization $Z^{(m)}$ of
some smooth field configuration with topological charge $m \not= 0$.

Given the relation ${\rm Im} S_{\rm cont} [ Z^{(m)} , {\tilde Z}^{(m)} ] = 
2\pi m\sigma_{xy}^{(0)}$ and the requirement $S_{\rm cont} = S_{\rm latt}$ 
for smooth fields, the topological coupling is determined by
        \[
        {\rm Im} S_{\rm latt} [ Z^{(m)} , {\tilde Z}^{(m)} ]
        = 2\pi m \sigma_{xy}^{(0)} .
        \]
To calculate $\sigma_{xy}^{(0)}$ from this equation, it is easiest 
to consider fields that have topological charge $m = 1$ and are of the 
special form
        \[
        Z^{(1)} = \pmatrix{0 &0\cr 0 &f\cr} , \quad
        {\tilde Z}^{(1)} = \pmatrix{0 &0\cr 0 &-{\bar f}\cr} .
        \]
Here the components $Z_{\rm BB}$, $Z_{\rm BF}$, and $Z_{\rm FB}$, 
which are topologically trivial, have been set to zero and only
$Z_{\rm FF}$ has been retained.  With this choice, the formula 
for $\sigma_{xy}^{(0)}$ reduces to
        \begin{eqnarray}
        \sigma_{xy}^{(0)} &=& {1\over 2\pi} {\rm Im}
        S_{\rm latt} [ Z^{(1)} , {\tilde Z}^{(1)} ]
        = - {1\over 2\pi} {\rm Im} \ln {\rm Det}_{\cal H} 
        \left( 1 + {\bar f} U_1^{\vphantom{-1}} f U_1^{-1} \right)
        \nonumber \\
        &=& - {1\over 2\pi} \sum_{{\rm nodes} \ n} {\rm Im} \ln 
        {\cal R}_{\rm FF}
        \left( {\bar f}(U_n) , {\bar f}(D_n) , f(L_n) , f(R_n) \right) ,
        \nonumber
        \end{eqnarray}
where ${\cal R}_{\rm FF}$ is the R-matrix in the FF sector:
        \begin{eqnarray}        
        &&{\cal R}_{\rm FF} \left( 
        {\bar f}(U_n) , {\bar f}(D_n) , f(L_n) , f(R_n) \right) 
        \nonumber \\
        &=& {\rm Det} \pmatrix{ 1+{\bar f}(U)[f(L)+f(R)]/2
        &{\bar f}(U)[f(L)-f(R)]/2\cr {\bar f}(D)[f(L)-f(R)]/2
        &1+{\bar f}(D)[f(L)+f(R)]/2 \cr}
        \nonumber \\
        &=& 1 + {1\over 2}[{\bar f}(U) + {\bar f}(D)][f(L)+f(R)]
        + {\bar f}(U){\bar f}(D)f(L)f(R) .
        \nonumber
        \end{eqnarray}
Now consider in the 2d plane with coordinates $x$ and $y$ the smooth field 
configuration
        \[
        f(x,y) = \left( {x-y\over A_-} + i {x+y\over A_+} - z_0 \right)^{-1}
        \qquad (A_\pm \in {\Bbb R}) ,
        \]
which interpolates between $f = \infty$ (corresponding to the south
pole on the two-sphere $M_{\rm F} = {\rm S}^2$) at 
        \begin{equation}
        x+iy = {1-i\over 2} A_- {\rm Re}z_0 + {1+i\over 2} A_+ {\rm Im}z_0
        \label{top_singularity}
        \end{equation}
and $f = 0$ (the north pole on ${\rm S}^2$) at infinity, and has
topological charge $m = 1$, as can easily be checked.  (Note that $f$
is {\it not} a one-instanton configuration, unless $A_- = A_+$.)  The
parameters $A_\pm$ set the size in $x \pm y$ of the region where the
topological density $L_{\rm top}$ for $f$ is appreciably different from 
zero.

For this choice of $f$, we now work out the R-matrix at the node
in Fig.~3a with coordinates $(n_x, n_y)$.  Setting $\alpha_\pm = a / 
(2 A_\pm)$, and $\zeta = (n_x-n_y)a/A_- + i(n_x+n_y)a/A_+ - z_0$,
we have
        \begin{eqnarray}
        {\bar f}(U) &=& (\bar\zeta - \alpha_- - i \alpha_+)^{-1}, \quad
        {\bar f}(D) = (\bar\zeta + \alpha_- + i \alpha_+)^{-1},
        \nonumber \\
        f(L) &=& (\zeta - \alpha_- - i \alpha_+)^{-1}, \quad
        f(R) = (\zeta + \alpha_- + i \alpha_+)^{-1} .
        \nonumber
        \end{eqnarray}
The resulting value of the R-matrix is
        \[
        {\cal R}_{\rm FF} = 1 + {2|\zeta|^2 + 1\over [\zeta^2 
        - (\alpha_- + i\alpha_+)^2][{\bar\zeta}^2 - (\alpha_-
        + i\alpha_+)^2]} \ .
        \]
In order for the field configuration to be smooth on the lattice, the
parameters $\alpha_\pm$ must be small.  This allows us to Taylor expand 
with respect to one of these, say $\alpha_+$.  Taking also the imaginary 
part of the logarithm, we obtain
        \[
        {\rm Im} \ln {\cal R}_{\rm FF} = 4 \alpha_+ \alpha_- \rho(1+\rho)^{-1}
        {\rm Re} \left( \zeta^2 - \alpha_-^2 \right)^{-1} + 
        {\cal O}(\alpha_+^2) ,
        \]
where $1 + \rho = {\cal R}_{\rm FF}|_{\alpha_+ = 0}$.

In the next step we sum the contributions from all nodes $(n_x + k ,
n_y + k)$ with $k \in {\Bbb Z}$.  The smallness of $\alpha_+ \ll 1$
allows us to convert the expression for $\sum {\rm Im}\ln {\cal
  R}_{\rm FF}$ into an integral.  Decomposing $\zeta$ into real and
imaginary parts by $\zeta = \xi + i\eta$,
        \[
        \xi = 2\alpha_- (n_x - n_y) - {\rm Re}z_0 ,
        \quad
        \eta = 2\alpha_+ (n_x + n_y) - {\rm Im}z_0 ,
        \]
we arrive at
        \[
        \sum_k {\rm Im}\ln {\cal R}_{\rm FF}\Big|_{(n_x+k,n_y+k)}
        {\buildrel \alpha_+ \to 0 \over \longrightarrow}
        \alpha_- \int_{\Bbb R} d\eta {\rho \over 1 + \rho} \ {\rm Re}
        \left( (\xi+i\eta)^2 - \alpha_-^2 \right)^{-1} .
        \]
This integral is easily evaluated by closing the contour and applying
the method of residues.  One finds
        \[
        \alpha_- \int_{\Bbb R} d\eta {\rho \over 1 + \rho} \ {\rm Re}
        \left( (\xi+i\eta)^2 - \alpha_-^2 \right)^{-1}
        = \cases{-\pi, &if $|\xi| < \alpha_-$; \cr
        0, &otherwise. \cr}
        \]
        
What is the geometric interpretation of this result?  Let us agree
that the word ``vertex'' here means a node taken together with its four
links, severed at half the distance to neighboring nodes.  The set of
vertices with node coordinates $(n_x + k,n_y + k)$ $(k\in{\Bbb Z})$
sweep out a diagonal strip of the 2d network.  The above result means
that $\sum_k {\rm Im}\ln {\cal R}_{\rm FF}$ vanishes if the center
(\ref{top_singularity}) of the topological excitation lies outside the
diagonal strip swept out by the vertices $(n_x + k,n_y + k)$ (with
variable $k\in{\Bbb Z}$), and equals $-\pi$ when it lies inside.

The above calculation applies to the type of node shown in Fig.~3a,
which may be characterized by the condition $n_x + n_y \in 2{\Bbb Z}$,
say.  Doing the calculation for the other type of node ($n_x + n_y
\in 2{\Bbb Z}+1$, or Fig.~3b) gives the same result except for a 
change of sign $-\pi \to \pi$.  Now observe that the diagonal
strips swept out by the vertices $(n + k,-n + k)$ and $(n + k, 
-n + 1 + k)$ $(k , n\in{\Bbb Z})$ cover the plane completely without 
overlapping.  Therefore, 
        \[
        \exp \ i {\rm Im}S_{\rm latt} [ Z^{(1)} , {\tilde Z}^{(1)} ] = 
        \exp \sum_{{\rm all \ nodes}} - i {\rm Im} \ln{\cal R}_{\rm FF}
        = e^{\pm i\pi} = - 1,
        \]
independent of the location of the topological singularity.
By $\sigma_{xy}^{(0)} = (2\pi)^{-1} {\rm Im} S_{\rm latt}[Z^{(1)} ,
{\tilde Z}^{(1)}]$, this proves $\sigma_{xy}^{(0)} = 1/2$ (mod 1). 

\subsection{Generalization to N channels}
\label{sec:N_channels}

D.K.K. Lee and Chalker \cite{lee_chalker} introduced a generalization
of the network model that has {\it two} channels per link.  The
one-step time evolution operator of that model, $U^{(2)}$, is again a
product of factors: $U^{(2)} = U_1^{(2)} U_0^{(2)}$.  The second
factor is diagonal on links and associates with each link $l$ a $2\times
2$ matrix $U_0(l)$ drawn at random from the unitary group in two-dimensional
channel space, ${\rm U}(2)$.  The first factor describes the
deterministic scattering at the nodes.  Depending on the choice made
for this factor, the two-channel model applies to a spin degenerate
Landau level or electrons in a random magnetic field (the so-called
random flux problem).

In this subsection we consider an $N$-channel generalization of the
one- and two-channel network models, where random ${\rm U}(N)$
matrices are placed on the links.  The one-step time evolution
operator is written $U^{(N)} = U_1^{(N)} U_0^{(N)}$.  The two factors
describe the deterministic $N$-channel scattering at the nodes, and
the random ${\rm U}(N)$ directed propagation along links,
respectively.

The case of $N$ channels per link can be treated by a slight 
extension of our field theoretic formalism.  Such an extension 
is possible since the basic formula (\ref{HS}) was shown 
in \cite{mrz_circular} to generalize from ${\rm U}(1)$ to 
${\rm U}(N)$, as follows:
        \begin{eqnarray}
        &&\int_{{\rm U}(N)} dU \exp \left( \bar\psi_{+\sigma}^i U^{ij}
        \psi_{+\sigma}^j + \bar\psi_{-\tau}^j {\bar U}^{ij} 
        \psi_{-\tau}^i \right)
        \nonumber \\
        &=& \int_{\rm Efetov} D\mu_N ( Z, {\tilde Z} ) \exp \left( 
        \bar\psi_{+\sigma}^i Z_{\sigma\tau}^{\vphantom{i}} \psi_{-\tau}^i 
        + \bar\psi_{-\tau}^j {\tilde Z}_{\tau\sigma}^{\vphantom{j}} 
        \psi_{+\sigma}^j \right) ,
        \nonumber
        \end{eqnarray}
where $U \equiv U_0(l)$, and the link label $l$ was suppressed for
clarity.  The right-hand side differs from that of (\ref{HS}) only by
the channel index $i = 1, ..., N$ attached to $\psi, \bar\psi$, and
the different form of the weight function in the superintegration
measure:
        \[
        D\mu_N(Z,\tilde Z) := D(Z,\tilde Z) \ {\rm SDet}(1-Z\tilde Z)^N .
        \]
Using the above generalization of (\ref{HS}), we can reformulate the 
$N$-channel model as a supersymmetric lattice field theory with action
        \[
        S_{\rm latt}^{(N)}[Z,\tilde Z] = 
        \ln {\rm SDet}_{{\cal H}\otimes{\Bbb C}^N} 
        \left( 1 - U_1^{(N)} Z U_1^{(N)\dagger} \tilde Z \right)
        - N \ln {\rm SDet}_{\cal H} ( 1 - Z \tilde Z) .
        \]
The derivation exactly parallels that of Sec.~\ref{sec:susy}.
The first superdeterminant runs over the tensor product of superspace 
with link space (${\cal H}$) and channel space (${\Bbb C}^N$).  

If we choose the deterministic scattering $U_1^{(N)}$ to be of the
special form $U_1^{(N)} = U_1^{(N=1)} \otimes 1_N$, i.e. $U_1^{(N)}$
acts as the identity in channel space, $S_{\rm latt}^{(N)}$ is simply
a multiple of the action of the one-channel model:
        \[
        S_{\rm latt}^{(N)} [Z,\tilde Z] = 
        N \times S_{\rm latt}^{(N=1)} [Z,\tilde Z] .
        \]
Therefore, from Secs.~\ref{sec:sig_xx} and \ref{sec:sig_xy} the 
coupling constants of the corresponding continuum field theory are
        \[
        \sigma_{xx}^{(0)} = \sigma_{xy}^{(0)} = N / 2 .
        \]
The choice $U_1^{(N)} = U_1^{(N=1)} \otimes 1_N$ for $N = 2$ is
appropriate \cite{lee_chalker} for the spin degenerate Landau level at
the center of the band, and for the random flux problem at the
symmetric point.  Because Pruisken's model at $\sigma_{xy}^{(0)} = 2 /
2 \ ({\rm mod} \ 1) = 0$ is known to be a massive theory, our result
for $\sigma_{xy}^{(0)}$ confirms the proposition
of \cite{lee_chalker} that neither of these systems is critical.
Rather, they are in a (Haldane type) massive phase \cite{DHLee},
corresponding to localization of all states.

\section{Symmetries of the R-matrix}
\label{sec:symmetries}

The basic building block of the SUSY reformulated Chalker-Coddington
model is its R-matrix.  For the case $N = 1$, which is to be
analysed here in more detail, this building block was given in
(\ref{R_matrix}).  To gain a deeper understanding of the model and
possible variations thereof, we are now going to investigate the
symmetries of that R-matrix.  Most of the effort will be expended
on rewriting the expression (\ref{R_matrix}) in a different form,
so as to make those symmetries more evident. 

We start by undoing the $\psi,\bar\psi$ integration to write the
R-matrix as 
        \begin{eqnarray}
        &&{\cal R} = \prod_{I,O} 
        {\rm SDet}\left( 1 - {\tilde Z}(O) Z(O) \right)^{1/2}
        {\rm SDet}\left( 1 - {\tilde Z}(I) Z(I)\right)^{1/2}
        \nonumber \\
        &&\hspace{2cm}
        \times \int D(\psi,\bar\psi) \exp \Big( - \bar\psi_{+\sigma}(O)
        \psi_{+\sigma}(O) 
        + \bar\psi_{+\sigma}(O) U_1(O,I) Z_{\sigma\tau}(I) \psi_{-\tau}(I)
        \nonumber \\
        &&\hspace{5.5cm}
        - \bar\psi_{-\tau}(I) \psi_{-\tau}(I) 
        + \bar\psi_{-\tau}(I) U_1^{-1}(I,O) {\tilde Z}_{\tau\sigma}(O) 
        \psi_{+\sigma}(O)
        \Big) .
        \nonumber
        \end{eqnarray}
Here we have introduced the labels $I \in \{ i1,i2\} := \{ L,R\}$ and
$O\in \{ o1,o2\} := \{ U,D\}$.  The notation is motivated by observing
that, according to the direction of motion indicated by the arrows in
Fig.~3, the links $L$ and $R$ are {\it incoming} states for the
scattering at the nodes, while the links $U$ and $D$ are {\it outgoing} 
states.

To proceed, we need to recall briefly various mathematical structures
that were developed in detail in the appendices of \cite{mrz_circular}
and \cite{spinchain}.  First of all, we introduce Fock operators $c$
and $\bar c$, which are quantum counterparts of the classical
variables $\psi$ and $\bar\psi$.  Let $b_\pm^\dagger ,
b_\pm^{\vphantom{\dagger}}$ and $f_\pm^\dagger ,
f_\pm^{\vphantom{\dagger}}$ be canonical boson and fermion
creation and annihilation operators, and set
        \begin{eqnarray}
        c_{+{\rm F}} &=& f_+ , \quad c_{+{\rm B}} = b_+ , \quad
        c_{-{\rm F}} = f_-^\dagger , \quad c_{-{\rm B}} = b_-^\dagger ,
        \nonumber \\
        {\bar c}_{+{\rm F}} &=& f_+^\dagger , \quad {\bar c}_{+{\rm B}} 
        = b_+^\dagger , \quad
        {\bar c}_{-{\rm F}} = f_-, \quad {\bar c}_{-{\rm B}} = - b_- .
        \nonumber 
        \end{eqnarray}
The operators $c$ and $\bar c$ are canonical pairs, with graded (or super)
commutation relations
        \[
        [ c_X , {\bar c}_Y ] := c_X {\bar c}_Y - (-1)^{|X||Y|} 
        {\bar c}_Y c_X = \delta_{XY} ,
        \]
where $|X| = 0$ if $X = \pm{\rm B}$ and $|X| = 1$ if $X = \pm{\rm F}$.  
They act in a Bose-Fermi Fock space with vacuum
        \begin{equation}
        c_{+{\rm B}} | 0 \rangle = c_{+{\rm F}} | 0 \rangle = 
        {\bar c}_{-{\rm B}} | 0 \rangle = {\bar c}_{-{\rm F}} | 0 \rangle 
        = 0 .
        \label{vacuum_even}
        \end{equation}
The graded commutation relations are invariant under canonical
transformations, ${\bar c}_X \mapsto T_g^{\vphantom{-1}} {\bar
  c}_X^{\vphantom{-1}} T_g^{-1} = {\bar c}_Y g_{YX}$ and $c_X \mapsto
T_g^{\vphantom{-1}} c_X^{\vphantom{-1}} T_g^{-1} = (g^{-1})_{XY}
c_{Y}$, where
        \[
        g = \pmatrix{A &B\cr C &D\cr} \in {\rm GL}(2|2) \mapsto T_g := 
        \exp \left( {\bar c}_X (\ln g)_{XY} c_Y \right)
        \]
defines a representation of ${\rm GL}(2|2)$ on Fock space. 

Consider now the subspace, $V$, selected by the condition ${\bar c}_X
c_X = 0$ (summation convention!) or, equivalently, by
        \[
        b_+^\dagger b_+^{\vphantom{\dagger}} + 
        f_+^\dagger f_+^{\vphantom{\dagger}} = 
        b_-^\dagger b_-^{\vphantom{\dagger}} +
        f_-^\dagger f_-^{\vphantom{\dagger}} .
        \]
The resulting constraint on the Bose-Fermi occupation numbers is
$n_{b+} + n_{f+} = n_{b-} + n_{f-}$.  By using this equation to
eliminate $n_{b-}$, say, we can characterize the states of $V$ by
a triplet of integers $(n_{b+},n_{f+},n_{f-})$, where $n_{f+}$ and
$n_{f-}$ take values from the set $\{0,1\}$, and $n_{b+} = 0,1,2,...,
\infty$.

Let $P_V$ be the operator that projects Fock space onto the
subspace $V$.  As was shown in \cite{mrz_circular}, this projector 
has a resolution
        \[
        P_V = \int D(Z,\tilde Z) \ |Z\rangle \langle Z| 
        \]
by generalized coherent states,
        \begin{eqnarray}
        | Z \rangle &=& \exp \left( {\bar c}_{+\sigma} Z_{\sigma\tau}
        c_{-\tau} \right) | 0 \rangle \ {\rm SDet} (1 - Z\tilde Z)^{1/2} ,
        \nonumber \\
        \langle Z | &=& {\rm SDet}(1 - Z\tilde Z)^{1/2} \ \langle 0 |
        \exp \left( - {\bar c}_{-\tau} {\tilde Z}_{\tau\sigma} 
        c_{+\sigma} \right) .
        \nonumber
        \end{eqnarray}

By simple manipulations exploiting the standard properties of coherent 
states, one can verify the following equality:
        \begin{eqnarray}
        &&\prod_I {\rm SDet} \left( 1 - {\tilde Z}(I) Z(I) \right)^{1/2}
        \times \exp \left( \bar\psi_{+\sigma}(O) U_1(O,I) Z_{\sigma\tau}(I) 
        \psi_{-\tau}(I) \right) 
        \nonumber \\
        &=& \langle 0 | \exp \left( 
        \bar\psi_{+\sigma}(o\mu) U_1(o\mu , i\nu) c_{+\sigma}(\nu) - 
        {\bar c}_{-\tau}(\nu) \psi_{-\tau}(i\nu) \right) | Z(i1) \otimes 
        Z(i2) \rangle
        \nonumber \\
        &=& \langle 0 | \exp \left( \bar\psi_{+\sigma}(o\mu) 
        c_{+\sigma}(\mu) - {\bar c}_{-\tau}(\nu) \psi_{-\tau}(i\nu) \right)
        \nonumber \\
        &&\hspace{3cm}\times \exp \left(
        {\bar c}_{+\sigma}(\mu) (\ln \hat U_1)(\mu,\nu) c_{+\sigma}(\nu) 
        \right) | Z(i1) \otimes Z(i2) \rangle .
        \nonumber
        \end{eqnarray}
Here $\mu , \nu \in \{ 1,2\}$, and $\hat U_1$ is defined by identifying 
initial and final channels, i.e. $\hat U_1(\mu,\nu) = U_1(o\mu,i\nu)$.  
Similarly,
        \begin{eqnarray}
        &&\prod_O {\rm SDet} \left( 1 - {\tilde Z}(O) Z(O) \right)^{1/2}
        \times \exp \left( {\bar\psi}_{-\tau}(I) U_1^{-1}(I,O) 
        {\tilde Z}_{\tau\sigma}(O) \psi_{+\sigma}(O) \right) 
        \nonumber \\
        &=& \langle Z(o1) \otimes Z(o2) | \exp \left( 
        {\bar\psi}_{-\tau}(i\nu) U_1^{-1}(i\nu,o\mu) c_{-\tau}(\mu)
        + {\bar c}_{+\sigma}(\mu) \psi_{+\sigma}(o\mu) \right) | 0 \rangle
        \nonumber \\
        &=& \langle Z(o1) \otimes Z(o2) | \exp \left( {\bar c}_{-\tau}(\nu) 
        (\ln \hat U_1)(\nu,\mu) c_{-\tau}(\mu) \right) 
        \nonumber \\
        &&\hspace{5cm}\times\exp\left(
        {\bar c}_{+\sigma}(\mu) \psi_{+\sigma}(o\mu) + 
        {\bar\psi}_{-\tau}(i\nu) c_{-\tau}(\nu) \right) | 0 \rangle .
        \nonumber
        \end{eqnarray}
The variables $\psi, \bar\psi$ have now served their purpose and we
integrate them out, by using the closure relation for Bose-Fermi 
coherent states:
        \begin{eqnarray}
        {\rm id} &=& \int D(\psi,\bar\psi) \exp \left(
        - \bar\psi_{+\sigma}(\mu) \psi_{+\sigma}(\mu) 
        - \bar\psi_{-\tau}(\nu) \psi_{-\tau}(\nu) \right)
        \nonumber \\
        &&\times 
        \exp \left( {\bar c}_{+\sigma}(\mu)\psi_{+\sigma}(\mu) 
        + \bar\psi_{-\tau}(\nu) c_{-\tau}(\nu) \right) | 0 \rangle \langle 0 |
        \exp \left( \bar\psi_{+\sigma}(\mu)c_{+\sigma}(\mu) - 
        {\bar c}_{-\tau}(\nu) \psi_{-\tau}(\nu) \right) .
        \nonumber
        \end{eqnarray}
All this results in the following formula for the R-matrix:
        \[
        {\cal R}\left( {\tilde Z}(U) , {\tilde Z}(D), Z(L) , Z(R) \right) =
        \langle Z(U) \otimes Z(D) | \hat{\cal R} | Z(L) \otimes Z(R) \rangle ,
        \]
where the operator $\hat{\cal R}$ is expressed by
        \begin{equation}
        \hat{\cal R} = \exp \left( 
        {\bar c}_X(\mu) (\ln {\hat U}_1)(\mu,\nu) c_X(\nu) \right) .
        \label{R_operator}
        \end{equation}
The advantage of this reformulation is that the invariance of $\hat{\cal R}$
under ${\rm GL}(2|2)$ transformations
        \begin{eqnarray}
        {\bar c}_X(\mu) \mapsto T_g^{\vphantom{-1}} 
        {\bar c}_X(\mu)^{\vphantom{-1}} T_g^{-1} &=& {\bar c}_Y(\mu) g_{YX} , 
        \nonumber \\
        c_X(\mu) \mapsto T_g^{\vphantom{-1}} c_X(\mu)^{\vphantom{-1}} 
        T_g^{-1} &=& (g^{-1})_{XY} c_Y(\mu)
        \nonumber
        \end{eqnarray}
is obvious whereas previously, in Sec.~\ref{sec:vertex}, we had to
work quite hard to establish the global ${\rm GL}(2|2)$ invariance. 

Although $\hat{\cal R}$ commutes with $T_g$, its matrix elements are
{\it not} invariant:
        \[
        {\cal R}\left( {\tilde Z}(U) , {\tilde Z}(D), Z(L) , Z(R) \right) 
        \not= {\cal R}\left( g\cdot{\tilde Z}(U) , g\cdot{\tilde Z}(D), 
        g\cdot Z(L) , g\cdot Z(R) \right) .
        \]
It is worth spending a little effort to explain how that comes about.
As will be seen, the reason is that, since the Fock space vacuum is
not a scalar with respect to $H = {\rm GL}(1|1) \times {\rm GL}(1|1)$,
the coherent states do not transform as functions on $G / H$, but
rather as sections of an associated line bundle,
see \cite{mrz_iqhe} and references therein.  In other words,
        \[
        T_g | Z \rangle \not= | g \cdot Z \rangle .
        \]
The correct transformation law is derived as follows.  We
write the coherent states as $|Z\rangle = T_{s(Z,\tilde Z)} |0\rangle$,
where
        \begin{eqnarray}
        s(Z,\tilde Z) &=& 
        \pmatrix{ (1-Z\tilde Z)^{-1/2} &Z(1-\tilde Z Z)^{-1/2}\cr
        {\tilde Z}(1-Z\tilde Z)^{-1/2} &(1-{\tilde Z}Z)^{-1/2}\cr}
        \nonumber \\
        &=& \pmatrix{1 &Z\cr 0 &1\cr} \pmatrix{(1-Z\tilde Z)^{+1/2} &0\cr
        0 &(1-\tilde Z Z)^{-1/2} \cr} \pmatrix{1 &0\cr \tilde Z &1\cr} .
        \nonumber
        \end{eqnarray}
We then define an $H$-valued function $h(g;Z,\tilde Z)$ by
        \[
        g \ s(Z,\tilde Z) = s(g\cdot Z,g\cdot\tilde Z) \ h(g;Z,\tilde Z) .
        \]
The explicit form of $h(g;Z,\tilde Z)$ can be found in Appendix B
of \cite{mrz_iqhe}.  {}From (\ref{vacuum_even}) and the definition $T_g =
\exp \left( {\bar c}_X (\ln g)_{XY} c_Y \right)$, one easily sees that 
the vacuum carries a one-dimensional representation $\mu$ of $H$:
        \[
        T_h | 0 \rangle = | 0 \rangle \ \mu(h) \quad ({\rm for} \ h\in H),
        \qquad \mu \left( {\rm diag}(A,D) \right) = {\rm SDet}D^{-1} .
        \]
Therefore, the coherent states transform as
        \[
        T_g | Z \rangle = T_g T_{s(Z,\tilde Z)} | 0 \rangle = 
        T_{s(g\cdot Z,g\cdot\tilde Z)} T_{h(g;Z,\tilde Z)}
        | 0 \rangle = | g\cdot Z\rangle \ \mu\left( h(g;Z,\tilde Z) \right).
        \]
As a result, the R-matrix obeys the following transformation law:
        \begin{eqnarray}
        &&{\cal R}\left( g\cdot{\tilde Z}(o1) , g\cdot{\tilde Z}(o2), 
        g\cdot Z(i1) , g\cdot Z(i2) \right) = 
        {\cal R}\left({\tilde Z}(o1),{\tilde Z}(o2),Z(i1),Z(i2)\right)
        \nonumber \\
        &&\hspace{3.5cm}\times
        \prod_{\lambda=1,2} \mu\left(h(g;Z(o\lambda),\tilde Z(o\lambda))\right)
        \prod_{\nu=1,2} \mu\left(h(g;Z(i\nu),\tilde Z(i\nu))\right)^{-1} .
        \nonumber
        \end{eqnarray}
Because each link is incoming with respect to one node, and outgoing
with respect to another, the multipliers $\mu\left(h(g;Z,\tilde
  Z)\right)$ cancel when all R-matrices are multiplied together, so
that the global ${\rm GL}(2|2)$ invariance is recovered.

The transformation law for the R-matrix explains why a direct
gradient expansion to extract $\sigma_{xy}^{(0)}$ from $S_{\rm latt}$
is difficult.  Such an expansion locally produces terms such as
        \[
        {\rm STr} (1-{\tilde Z}Z)^{-1} {\tilde Z}\partial_x Z
                  (1-{\tilde Z}Z)^{-1} {\tilde Z}\partial_y Z ,
        \]
for example, which are {\it singular} at $Z_{\rm FF} = \infty$ (the
south pole on $M_{\rm F} = {\rm S}^2$).  When all of these terms are
correctly summed over the entire network, they cancel, as is
guaranteed by the global ${\rm GL}(2|2)$ invariance, which permits to
rotate the south pole into any other point on the two-sphere ${\rm
  S}^2$.  However, the cancellation really does take place only {\it
  after} summation of terms.  By the multiplier-corrected
transformation law of the R-matrix, singular terms remain locally,
making the extraction of the topological coupling difficult.  This,
then, is the reason why a gradient expansion was not attempted in
Sec.~\ref{sec:sig_xy}.
 
\section{Anisotropic Limit}
\label{sec:anisotropic}

We have presented an analytical method for dealing with the
Chalker-Coddington model in its original {\it isotropic} formulation,
by mapping it on a lattice equivalent of Pruisken's nonlinear $\sigma$
model.  In this section we will review another way of arriving at
Pruisken's model, a replica version of which was first published by
D.H. Lee.  Following \cite{DHLee} we now take for our starting point
the {\it anisotropic} limit of the Chalker-Coddington model, and
replace the unitary operator $U = U_1 U_0$ by the Hamiltonian $H$ for
an array of chiral modes $n = 1, 2, ...$ with velocity $v$ and
alternating direction of propagation:
        \[
        H = \oint dx \sum_{n,n'} \Psi_n^\dagger(x) 
        \left[ \delta_{nn'} (-1)^n iv\partial_x + V_{nn'}(x) 
        \right] \Psi_{n'}(x) .
        \]
The functions $V_{nn'}(x) = {\bar V}_{n'n}(x)$ are uncorrelated 
Gaussian random variables with zero mean and variance
        \[
        \langle V_{nn'}(x) {\bar V}_{nn'}(x') \rangle = 
        2 \left( u_0 \delta_{nn'} + u_1 \delta_{n,n'+1} + u_1
        \delta_{n,n'-1} \right) \delta(x-x') .
        \]
The symbol $\oint$ means that we are using periodic boundary 
conditions in $x$. 

To prepare the treatment of the general case, we shall first consider 
the case of a single chiral mode $n$.  The supersymmetric generating
functional for the retarded and advanced Green's functions of $H$ is 
set up in the usual way, see Sec.~\ref{sec:susy}.  Ensemble averaging
over the random potential $V(x) = V_{nn}(x)$ leads to the functional
integral
       \[
       {\cal Z} = \int {\cal D}(\psi,\bar\psi) \ 
       \exp \oint dx \left( \bar\psi (\Lambda v\partial_x - \varepsilon)
       \psi - u_0 (\bar\psi \Lambda \psi)^2 \right) ,
       \]
where $\varepsilon$ is a positive infinitesimal.
As before, $\psi_X(x)$ is a super ``spinor'' field with four components
$X = +{\rm B}$ (retarded Boson), $X = +{\rm F}$ (retarded
Fermion), $X = -{\rm B}$ (advanced Boson), and $X = -{\rm
  F}$ (advanced Fermion).  The notation means $\bar\psi\Lambda\psi = 
\bar\psi_{+\sigma}\psi_{+\sigma} - \bar\psi_{-\tau}\psi_{-\tau}$.
If the energy in the retarded and advanced sectors is different,
$\omega = E_+ - E_- \not= 0$, we need to add a term $i\omega\bar\psi\psi$
to the Lagrangian.  To probe this field theory more generally, we may 
couple it to an external nonabelian gauge field $A(x) \in {\rm Lie}
\left( {\rm GL}(2|2) \right)$ and consider
       \[
       {\cal Z}_{\rm Dirac}^{u} [A] := \int {\cal D}(\psi,\bar\psi) \ 
       \exp \oint dx \left( \bar\psi \Lambda v ( \partial_x + A ) \psi
       - u (\bar\psi \Lambda \psi)^2 \right) .
       \]
We have temporarily set $u = u_0$ for notational simplicity.  The
special coupling to frequency is retrieved by putting $A = (i\omega -
\varepsilon)\Lambda/v$, independent of $x$.  Note that the generating
functional ${\cal Z}_{\rm Dirac}^{u}[A]$ is invariant under local
gauge transformations,
       \[
       {\cal Z}_{\rm Dirac}^{u}[A] = {\cal Z}_{\rm Dirac}^{u}[{}^h A], 
       \quad {}^hA = hAh^{-1} + h \partial_x h^{-1} ,
       \]
where $h(x) \in {\rm GL}(2|2)$ acts on the spinor field by $\psi 
\mapsto h \psi$ and $\bar\psi\Lambda \mapsto \bar\psi\Lambda h^{-1}$.

It turns out that one can write down another $0 + 1$ dimensional field
theory that has the very same local gauge invariance.  The field of
this theory is the supermatrix $Q = g\Lambda g^{-1}$, which was
defined in Sec.~\ref{sec:pruisken} and transforms as $Q \mapsto
h Q h^{-1}$.  The generating functional is
       \[
       {\cal Z}_{\rm WZ}[A] := \int {\cal D}Q \ \exp \oint dx \
       {\rm STr} {\Lambda\over 2} g^{-1} (\partial_x + A)g .
       \]
Clearly, this satisfies ${\cal Z}_{\rm WZ}[A] = {\cal Z}_{\rm WZ}[{}^h
A]$.  The linear derivative term in the action is of the Wess-Zumino
type, i.e. cannot be expressed in a globally nonsingular way in terms
of $Q$, and is often called a Berry phase.  The theory is well-defined
because the ambiguity under right translations $g \mapsto g h_{\rm R}$
[$h_{\rm R}^{\vphantom{-1}}\Lambda h_{\rm R}^{-1} = \Lambda$ or,
equivalently, $h_{\rm R}(x)\in{\rm GL}(1|1)\times{\rm GL}(1|1)$] gives
rise to a factor
       \[
       \exp \oint dx {\rm STr}{\Lambda\over 2} h_{\rm R}^{-1}
       \partial_x h_{\rm R} = 
       \exp \oint dx \partial_x {\rm STr}{\Lambda\over 2} \ln h_{\rm R}
       = \exp 2\pi im = 1 ,
       \]
which is unobservable in the functional integral. (For more details
see Sec.~3.4 of \cite{mrz_iqhe}.)

The local gauge invariance shared by ${\cal Z}_{\rm Dirac}^{u}[A]$ and 
${\cal Z}_{\rm WZ}[A]$ suggests the existence of some relation between
these theories.  In fact, the following statement is true:
       \begin{equation}
       \lim_{u \to\infty} {\cal Z}_{\rm Dirac}^{u}[A] = 
       {\cal Z}_{\rm WZ}[A] .
       \label{nonabelian_bosonization}
       \end{equation}
This identity can be viewed as a $0+1$ dimensional analog of nonabelian
bosonization in $1+1$ dimensions and, since $A$ couples to
$v\psi\bar\psi\Lambda$ and $g\Lambda g^{-1}/2 = Q/2$ in the respective
cases, amounts to the ``bosonization rule''
       \[
       v \psi\bar\psi\Lambda \ {\buildrel u \to\infty\over\longrightarrow} 
       \ Q/2 .
       \]

We now briefly sketch the proof of the nonabelian bosonization formula
(\ref{nonabelian_bosonization}).  By the local gauge invariance of
both theories, it is sufficient to prove the equality for an
$x$-independent gauge field $A$.  Moreover, $A$ may be taken to be a
diagonal matrix.  In this special case, it is easy to apply a method
that was described at length in \cite{spinchain} and works as follows.
As a first step, one identifies ${\cal Z}_{\rm Dirac}^{u}[A]$ as the
coherent state path integral of a supersymmetric Hubbard Hamiltonian
for bosons and fermions.  Then, one takes advantage of the limit $u
\to \infty$, which enforces a Hubbard constraint reducing the (low
energy) degrees of freedom to that of a single superspin.  And
finally, one sets up the coherent state path integral for the
superspin Hamiltonian.  The latter path integral turns out to be
${\cal Z}_{\rm WZ}[A]$, which completes the proof.

This proof, although straightforward, has the disadvantage of being
somewhat indirect.  A more direct procedure is to decouple the
interaction term $(\bar\psi\Lambda\psi)^2$ by introducing a
Hubbard-Stratonovitch field $Q$ coupling to $\psi\bar\psi\Lambda$ and
then to integrate out $\psi$ and $\bar\psi$.  The effective action
for $Q$ is
       \[
       S[Q] =  \oint dx \ {\rm STr} \left( -{u\over 4v^2} Q^2
       + \ln(\partial_x + A + u Q / v^2) \right) .
       \]
The next step is to simplify the $Q$ field functional integral by
means of the saddle-point approximation, as a result of which $Q$ gets
restricted to the nonlinear space $Q = g \Lambda g^{-1}$.  This step,
while only approximate in general, here becomes {\it exact} in the
limit $u \to\infty$.  [What makes this possible is the {\it
  stationarity} of the average density of states of $v i\partial_x +
V(x)$.]  By expanding $\ln \left( \Lambda g^{-1}(\partial_x + A)g
  +u/v^2 \right)$ to linear order in $g^{-1}(\partial_x + A)g$, one
obtains the action of the Wess-Zumino functional ${\cal Z}_{\rm
  WZ}[A]$.  Higher orders are suppressed by powers of $v^2/(L_x u)$,
with $L_x$ the system size.

Let us finally return to the case of many counterpropagating chiral
modes that are coupled by hopping matrix elements between neighboring
modes, with variance $u_1$.  The Gaussian random hopping gives rise to
an additional term in the Lagrangian,
       \begin{eqnarray}
       L \mapsto &&L + 2u_1 \sum_n \left( {\bar\psi}_n \Lambda \psi_{n+1} 
       \right) \left( {\bar\psi_{n+1}} \Lambda \psi_n \right) 
       \nonumber \\
       = &&L + 2u_1 \sum_n {\rm STr} \left( \psi_n \bar\psi_n \Lambda \right)
       \left( \psi_{n+1}\bar\psi_{n+1} \Lambda \right) .
       \nonumber
       \end{eqnarray}
By the bosonization rule $v \psi{\bar\psi}\Lambda \to Q/2$ for $u_0
\to\infty$, the additional term turns into $(u_1/2v^2) \sum_n {\rm STr}
(Q_n Q_{n+1})$ .  The condition of validity of this step is $u_0 \gg
u_1$.  As a result we obtain the $Q$ field action
       \[
       S[Q] = \oint dx \sum_n {\rm STr} \left( 
       (-1)^n {\Lambda\over 2} g_n^{-1}\partial_x g_n^{\vphantom{-1}}
       + {u_1\over 2v^2} Q_n Q_{n+1} \right) .
       \]
By a standard calculation \cite{fradkin,spinchain} this is the action
of the coherent state path integral for a quantum superspin Hamiltonian,
       \begin{equation}
       H_{\rm spin} = {2u_1\over v^2} \sum_n \sum_{XY} (-1)^{|Y|+1}
       S_n^{XY} S_{n+1}^{YX} ,
       \label{spin_Hamiltonian}
       \end{equation}
where $S^{XY} = {\bar c}_X c_Y$, and the graded commutation relations
of the Fock operators $c, {\bar c}$ were given in Sec.~\ref{sec:symmetries}.
To reproduce the alternating sign of the Wess-Zumino term, we must
alternate the definition of the Fock vacuum.  On even sites
$(n \in 2{\Bbb Z})$ the relations (\ref{vacuum_even}) apply, whereas on odd
sites $(n \in 2{\Bbb Z}+1)$ we have
       \begin{equation}
       {\bar c}_{+{\rm B}} | 0 \rangle = {\bar c}_{+{\rm F}} | 0 \rangle = 
       {c}_{-{\rm B}} | 0 \rangle = {c}_{-{\rm F}} | 0 \rangle = 0 
       \label{vacuum_odd}
       \end{equation}
instead.  The alternating vacuum plays the same role as the N\'eel
state for ordinary spin systems and means that the superspin chain
is ``antiferromagnetic'' in character.

The Hamiltonian $H_{\rm spin}$ with translational invariant coupling
$J = 2u_1/v^2 > 0$ was shown in \cite{mrz_iqhe} to represent the low
energy limit of the quantum Hamiltonian of Pruisken's nonlinear
$\sigma$ model at criticality.  Thus, we finally conclude that the
anisotropic Chalker-Coddington model with homogeneous (on average)
inter-channel hopping is in the same universality class as Pruisken's
model at $\sigma_{xy}^{(0)} = 1/2$.  (Clearly, this line of reasoning
is much less direct than that presented for the isotropic model in
Secs.~\ref{sec:susy}--\ref{sec:continuum}.)

To conclude this section let me mention that the nonabelian
bosonization formula (\ref{nonabelian_bosonization}) extends to
$N$ channels:
       \begin{eqnarray}
       &&\int {\cal D}Q \ \exp \oint dx \ N
       {\rm STr} {\Lambda\over 2} g^{-1} (\partial_x + A) g 
       \nonumber \\
       &=& \lim_{u\to\infty} \int {\cal D}(\psi,\bar\psi) \ 
       \exp \oint dx \left( \bar\psi_n \Lambda ( \partial_x + A ) \psi_n
       - u (\bar\psi_n \Lambda \psi_{n'})(\bar\psi_{n'} \Lambda \psi_n)
       \right) .
       \nonumber
       \end{eqnarray}
The strategy of the proof is the same as for $N = 1$.  Using this
formula we can easily reproduce the $N$-channel result
$\sigma_{xy}^{(0)} = N / 2$ of Sec.~\ref{sec:N_channels}.  The
$N$-channel bosonization formula offers also a quick way of analysing
the zero-dimensional limit of the 2d chiral metal \cite{spinchain}.

\section{Modified network model}
\label{sec:integrable}

The primary goal of all field theoretic analysis of the
plateau-to-plateau transition in integer quantum Hall systems must be
to {\it identify the fixed point theory} that describes this
transition and uncover the {\it conformal structure} it is expected to
have.  Recent attempts \cite{MIT} in this direction started from the
observation that Pruisken's model or, rather, the closely related
Dirac theory with random mass, random scalar potential, and random
gauge field, has a global ${\rm GL}(2|2)$ symmetry for the case of one
retarded and one advanced Green's function.  (A finite imaginary part
of the energy argument of the Green's functions reduces this symmetry
to ${\rm GL}(1|1)\times{\rm GL}(1|1)$.  However,
Secs.~\ref{sec:cc_model} and \ref{sec:susy} show how such a symmetry
breaking can be avoided by calculating a conductance between interior
contacts of the network model.)  This symmetry was then assumed to be
promoted in the infrared to a Kac-Moody symmetry, which severely
restricts the number of possible candidates for the fixed point
theory.  Unfortunately, these attempts have not been successful so
far.  What is needed as additional input to such considerations, which
are purely algebraic, is a firm understanding of the Hilbert space
structure, or the representations involved.  It is one of the aims of
the present work to contribute to such an understanding.

Following up on unpublished work by N. Read, it was argued
in \cite{mrz_iqhe} that the quantum Hamiltonian of the critical theory
should be a superspin Hamiltonian of the type (\ref{spin_Hamiltonian})
acting on a space of states built from alternating ${\rm GL}(2|2)$
modules
       \[
       ... \otimes V \otimes V^* \otimes V \otimes V^* \otimes ... ,
       \]
where $V$ and $V^*$ are generated by the action $g \mapsto T_g =
\exp\left( {\bar c}_X (\ln g)_{XY} c_Y\right)$ of ${\rm GL}(2|2)$ on
the vacua (\ref{vacuum_even}) and (\ref{vacuum_odd}) respectively.
(As follows from Sec.~4.4 of \cite{mrz_iqhe}, the elements of $V$ and
$V^*$ have an interpretation as holomorphic and antiholomorphic
sections of a line bundle associated to $G \to G/H$ by the
one-dimensional representation $\mu$ of $H$.  This permits the
construction of a nondegenerate pairing between $V$ and $V^*$, so that
these spaces can be viewed as being dual to each other, as suggested
by our notation.)  It is then natural to ask whether one might be able
to construct an {\it integrable} superspin Hamiltonian, offering the
possibility of an analytical and exact computation of critical
properties.  This question will now be addressed in the light of the
results of Secs.~\ref{sec:symmetries} and \ref{sec:anisotropic}.

Recall the ``functional'' vertex model presentation
(\ref{vertex},\ref{R_matrix}) of the Chalker-Coddington model,
       \begin{eqnarray}
       &&{\cal Z} = \int {\cal D}(Z,\tilde Z) \prod_{{\rm nodes}\ n} {R}
       \left( {\tilde Z}(u_n) , {\tilde Z}(d_n) , Z(l_n) , Z(r_n) \right) ,
       \nonumber \\
       &&{R}\left( {\tilde Z}(1) , {\tilde Z}(2) , Z(3) , Z(4) \right) = 
       \langle Z(1) \otimes Z(2) | \hat{R} | Z(3) \otimes Z(4) \rangle ,
       \nonumber \\
       &&\hat{R} = \exp \left( {\bar c}_X(\mu) (\ln {\hat U}_1)(\mu,\nu)
       c_X(\nu) \right) .
       \label{hat_R}
       \end{eqnarray}
(Note the change in the notation for links from capital to small letters.)
It is instructive to pass from the {\it integration} over fields
$Z,\tilde Z$ to a {\it summation} over discrete degrees of freedom, as
follows.  Every link $l$ emanates from one node, and ends at one node.
Therefore, each $Z(l)$ occurs once in the ``bra'' and once in the
``ket'' of some R-matrix.  Recall from Sec.~\ref{sec:symmetries} that
$P_V$ denotes the projector from Bose-Fermi Fock space onto the ${\rm
  GL}(2|2)$ module $V$, where the elements of $V$, referred to as
superspin states, are labeled by three quantum numbers $n_{f\pm} \in
\{ 0,1 \}$ and $n_{b+} = 0,1,2,...\infty$.  On using the closure
relation
       \[
       \int \prod_{{\rm links} \ l} D\left( Z(l) , {\tilde Z}(l) \right)
       \ \otimes_{l} | Z(l) \rangle \langle Z(l) | = 
       \otimes_{l} P_V(l) ,
       \]
the lattice functional integral over $Z,\tilde Z$ turns into a
partition sum over superspin configurations $\{ \alpha(l) \}$ [with
$\alpha(l) \in V(l)$] of a vertex model defined over the tensor
product space $\otimes_{l} V(l)$:
       \[
       {\cal Z} = \sum_{\{ \alpha(l) \} } (-1)^{N_F} \prod_{{\rm nodes} 
       \ n} {R}_{\alpha(u_n)\alpha(d_n),\alpha(l_n)\alpha(r_n)} \ ,
       \qquad
       {R}_{\alpha\beta,\gamma\delta} = \langle \alpha \otimes \beta
       | \hat{R} | \gamma \otimes \delta \rangle ,
       \]
where $(-1)^{N_F}$ is a sign factor due to supersymmetry (the
partition sum is a supertrace), and $\hat{R} : V(l)\otimes V(r)
\to V(u) \otimes V(d)$ is still given by (\ref{hat_R}).  For $p = 0$
(left turns only), $\hat{R}$ can be seen to be the identity map,
while for $p = 1$ (right turns only) we have $\hat{R} = {\cal
  P}$, where
       \[
       {\cal P} | \alpha \otimes \beta \rangle =
       (-1)^{|\alpha| |\beta|} | \beta \otimes \alpha \rangle
       \]
is the graded permutation operator.

Let me mention in passing that the presentation as a superspin
partition sum can also be obtained from the network model {\it
  directly} \cite{grs}, without passing through the intermediate stage
of a $Z$ field formulation.  

How is the model built from the vertex ${R}_{\alpha\beta,
  \gamma\delta}$ related to the superspin Hamiltonian
(\ref{spin_Hamiltonian})?  Given ${R}_{\alpha\beta, \gamma\delta}$ we
can set up the row-to-row transfer matrix, $T$, illustrated in
Fig.~4a, by summing over the superspin degrees of freedom that are
situated on the horizontal links.  (The meaning of the arrows in the
present context will become clear below.)  The corresponding
Hamiltonian, defined as the logarithm of the transfer matrix, is
nonlocal in general.  A local Hamiltonian results on making the
following modification of the isotropic network model.  On nodes with
coordinates $(n_x,n_y)\in{\Bbb Z}^2$ such that $n_x + n_y \in 2{\Bbb
  Z}$, the R-matrix is taken as it stands; but on the other half of
the nodes ($n_x + n_y \in 2{\Bbb Z}+1$), we replace the left-right
asymmetry parameter $p$ by its complement $1-p$.  (At the level of the
random network model, this is precisely what one does to arrive at the
anisotropic limit of Sec.~\ref{sec:anisotropic}.)  The row-to-row
transfer matrix of the resulting {\it anisotropic} vertex model has
the property, for $p = 0$, of translating the system by one lattice
unit (Fig.~4b).  By differentiating the logarithm of this transfer
matrix at $p = 0$, one gets a superspin Hamiltonian that couples only
nearest neighbors and is precisely the Hamiltonian $H_{\rm spin}$ of
(\ref{spin_Hamiltonian}).

\begin{figure}
\hspace{0.5cm}
\epsfxsize=15cm
\epsfbox{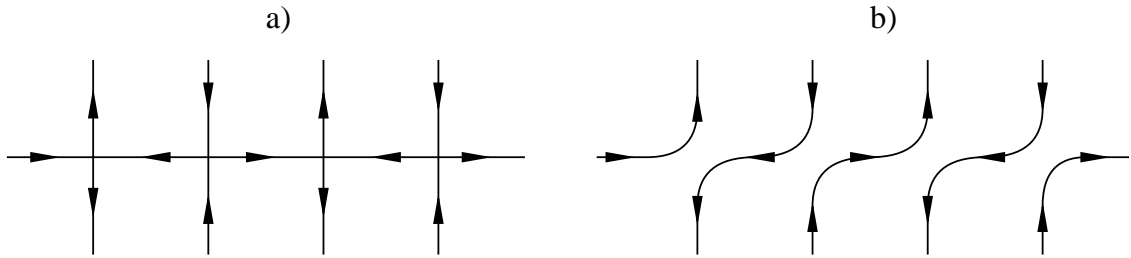}
\vspace{10pt}
\caption{a) Illustration of the row-to-row transfer matrix of the
  supersymmetric vertex model associated with the one-channel
  Chalker-Coddington model.  b) The transfer matrix of the anisotropic
  model at $p = 0$ translates the system by one lattice unit.}
\end{figure}

$H_{\rm spin}$ is not expected to be exactly solvable.  However, one
may ask whether it could be made so by slightly changing some
parameters while keeping the general structure and symmetries the
same.  To get some hint, we turn to the well-developed theory of
integrable systems \cite{baxter}.  There, the integrability of a 1d
quantum Hamiltonian $H$ is traced back to the existence of a transfer
matrix $T(u)$ that depends on a ``spectral'' parameter $u$ in such a
way that $T(u)T(v) = T(v)T(u)$ for all $u,v$, and $H$ is the
logarithmic derivative of $T(u)$ at some special point $u = u_0$.  A
sufficient condition for $T(u)$ to form a commuting family is known to
be the quantum Yang-Baxter equation for the R-matrix.

To my knowledge, most (if not all) of the integrable models discussed
in the literature have one characteristic feature in common: their
commuting family of transfer matrices possesses a ``classical'' limit,
where $T$ becomes the identity.  In contrast, for the row-to-row
transfer matrix associated to the Chalker-Coddington model (Fig.~4a),
no such limit exists.  The reason is simply this.  The R-matrix
$\hat{R}: V\otimes V \to V\otimes V$ in (\ref{R_matrix}) relates
incoming channels on horizontal (or vertical) links to outgoing
channels on vertical (or horizontal) links.  On the other hand, the
transfer matrix $T$ propagates the degrees of freedom from one row to
the next.  Therefore, to construct the row-to-row transfer matrix from
the R-matrix, we must reinterpret some initial states as final states,
and vice versa.  This is done by noting that the space of linear maps
$V \to V$ is isomorphic to the tensor product $V \otimes V^*$.  In
this way, one sees that in Fig.~4a vertical links with an arrow
pointing up carry the space $V$, whereas vertical links with an arrow
pointing down carry the dual space $V^*$.  (This, then, is the meaning
of the arrows in that figure.)  Hence, the row-to-row transfer matrix
of the SUSY reformulated Chalker-Coddington model is a map
       \[
       T : ... \otimes V \otimes V^* \otimes V \otimes ...
       \to ... \otimes V^* \otimes V \otimes V^* \otimes ... ,
       \]
which connects {\it inequivalent} spaces.  Therefore, $T$ cannot ever
be the identity.  This means that the Chalker-Coddington model lies
outside the category of vertex models for which the well-developed
Yang-Baxter machinery applies.  Thus it seems that there exists no
known systematic way of deforming the Chalker-Coddington model to
integrability and obtain an analytical solution.

The above discussion, though being a falsification, also suggests a
remedy.  Given that the standard formalism of the theory of integrable
systems requires the row-to-row transfer matrix to be a map
       \begin{equation}
       T : ... \otimes V \otimes V^* \otimes V \otimes ...
       \to ... \otimes V \otimes V^* \otimes V \otimes ... ,
       \label{transfer_matrix}
       \end{equation}
we can turn things around and modify the network model accordingly.
Consider the network shown in Fig.~5a.  As before, the electron
follows the direction of motion dictated by the arrows.  It picks up a
random ${\rm U}(1)$ phase while propagating along the links, and is
scattered deterministically at the nodes, just as for the original
one-channel Chalker-Coddington model.  The crucial difference from
Fig.~2 is that now the direction of motion {\it does not alternate but
  is invariant} along vertical and horizontal lines.  An electron
incident on a node either passes straight through it, with probability
$p$, say, or else is scattered to the right or left, as the case may
be, with probability $1-p$. (Note that such a modification of the
scattering dynamics has no justification from a microscopic picture of
guiding center drift along equipotentials.  However, since our aim is
only to describe the {\it critical} behavior, we should have a certain
amount of freedom in the choice of model.)

\begin{figure}
\hspace{0.5cm}
\epsfxsize=15cm
\epsfbox{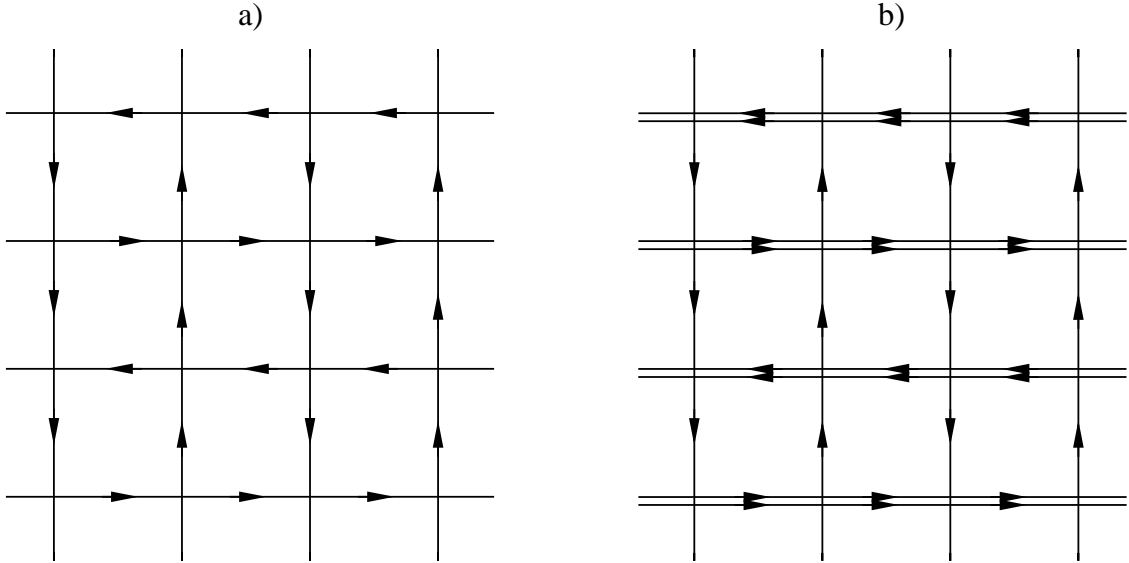}
\vspace{10pt}
\caption{a) One-channel Chalker-Coddington model, modified so as to make 
  the direction of motion invariant along every horizontal and vertical
  line.  b) The model of a) modified further, by doubling the number
  of channels on the horizontal links.}
\end{figure}

By construction, the row-to-row transfer matrix of the supersymmetric
vertex model associated with the network model of Fig.~5a is a map of
the desired type (\ref{transfer_matrix}).  One may now hope to be able
to deform and extend this transfer matrix to a one-parameter family of
commuting transfer matrices.  In somewhat more detail, this hope is
based on the following facts.  Consider the Lie superalgebra ${\cal
  G}$ of polynomial maps $u \in {\Bbb C} \to {\rm gl}(2,2)$ (with $u$
being the spectral parameter).  If $E_{AB}$ are the canonical
generators of ${\rm gl}(2,2)$, the classical $r$-matrix $r(u,v) =
(u-v)^{-1} \sum_{AB} E_{AB} (-1)^{|B|} E_{BA}$ gives rise to a
co(super)commutator ${\cal G} \to {\cal G}\otimes {\cal G}$ in the
usual way \cite{chari_pressley}, thereby turning ${\cal G}$ into a Lie
bisuperalgebra.  Quantization of ${\cal G}$ leads to a {\it Yangian
  superalgebra} ${\cal Y} := Y({\rm gl}(2,2))$, which is a ${\Bbb
  Z}_2$-graded Hopf algebra deformation of the universal enveloping
algebra of ${\rm gl}(2,2)$.  Let $\Delta : {\cal Y} \to {\cal
  Y}\otimes{\cal Y}$ be the comultiplication of ${\cal Y}$, and
$\Delta^{\rm op} = {\cal P} \circ \Delta$ its opposite.  According to
general principles \cite{drinfeld}, there exists a formal object called
the universal R-matrix ${\cal R} \in {\cal Y}\otimes{\cal Y}$, which
is determined by the intertwining relation $\Delta^{\rm op}(a) = {\cal
  R}\Delta(a){\cal R}^{-1}$ and has the expansion
       \[
       \ln {\cal R}(u) = u^{-1} \sum_{AB} (-1)^{|B|}
       E_{AB} \otimes E_{BA} + {\cal O}(u^{-3}) .
       \]
An irreducible matrix representation $\rho$ of ${\cal Y}$ yields
an $R$-matrix $R_\rho(u) = (\rho\otimes\rho)\bigl({\cal R}(u)\bigr)$
that is a rational function of $u$ and solves the quantum Yang-Baxter
equation.

When applying this formalism to our problem, we should beware of
potential problems due to the {\it infinite-dimensionality} of the
spaces $V$ and $V^*$.  Nevertheless, it does not seem unreasonable to 
expect the existence of R-matrices
       \begin{eqnarray}
       R(u) &=& (\rho_V \otimes \rho_V)\bigl({\cal R}(u)\bigr) \ : \
       V \otimes V \to V \otimes V ,
       \nonumber \\
       {\tilde R}(u) &=& (\rho_V \otimes \rho_{V^*})\bigl({\cal R}(u)\bigr) 
       \ : \ V \otimes V^* \to V \otimes V^* .
       \nonumber
       \end{eqnarray}
I have been able to verify that this expectation is fulfilled in the
case of $R(u)$, by explicit construction.  The existence of ${\tilde
  R}(u)$ remains an open question at the present time.  (Help from
experts on quantum groups would be very much appreciated.)  The latter
case is more complicated to treat because the decomposition of the
tensor product $V\otimes V^*$ into ${\rm gl}(2,2)$-irreducible
subspaces involves a continuous series of representations of ${\rm
  gl}(2,2)$ (see Sec. 5.2 of \cite{mrz_iqhe}), while in $V\otimes V$
only a discrete series appears.

Suppose now that both $R(u)$ and ${\tilde R}(u)$ exist, at least in
some domain of the spectral parameter $u$.  Then, since the quantum
Yang-Baxter equation is automatically fulfilled, we can build a
one-parameter family of commuting transfer matrices.  The next
question is: given the modifications we have made, is the physics of
such a model still that of the plateau transition?  Surely, obtaining
an exact solution remains a far goal.  It is therefore helpful that
the mapping onto Pruisken's nonlinear $\sigma$ model provides us with
a quick way to get oriented in the enlarged landscape of modified
network models.  Consider the model of Fig.~5a with parameter $p =
1/2$.  Taking the continuum limit and computing the topological
coupling in the same way as in Sec.~\ref{sec:sig_xy}, one finds
$\sigma_{xy}^{(0)} = 0 \ ({\rm mod} \ 1)$.  This means that the model
does {\it not} lie in the quantum Hall universality class, but is in a
massive phase with a finite (albeit large) localization length and
exponentially decaying correlations.

This result could have been anticipated from the following heuristic
argument.  Imagine separating the network into two independent
subsystems $A$ and $B$, one consisting of the vertical lines and the
other of the horizontal ones.  Then couple the modes within each
subsystem by weak tunneling amplitudes.  What you get in this way are
two copies of D.H. Lee's anisotropic limit of the Chalker-Coddington
model.  By the reasoning reviewed in Sec.~\ref{sec:anisotropic}, each
of these is critical, with the $\sigma$ model topological coupling
constants being $\sigma_{xy,A}^{(0)} = \sigma_{xy,B}^{(0)} = 1/2$.
Now join the two subsystems to form the network model of Fig.~5a.
{}From the meaning of $L_{\rm top}$ as a topological {\it density}, it
is reasonable to expect that $\sigma_{xy}^{(0)} = \sigma_{xy,A}^{(0)}
+ \sigma_{xy,B}^{(0)} = 2 \times 1/2 = 0 \ ({\rm mod} \ 1)$ for the
coupled system, if the fusion is done in such a way that
$\sigma_{xx}^{(0)}$ for the coupled system is spatially homogeneous.
Thus the network model of Fig.~5a will be noncritical at $p = 1/2$.
(Its correlation functions at large scales should be similar to those
of the two-channel Chalker-Coddington model at the symmetric point.)
Since noncriticality is a generic property, this will remain so in a
neighborhood of the point $p = 1/2$. 

In view of this heuristic argument, we expect that criticality can be
restored by superimposing on the noncritical network of Fig.~5a yet
{\it another} copy of the anisotropic network model, thereby producing
the network of Fig.~5b.  There, the horizontal lines carry two, rather
than one, channels per link.  The propagation on vertical links is
governed by random ${\rm U}(1)$ phases, as before, but the horizontal
links now carry random ${\rm U}(2)$ matrices (just like the
two-channel Chalker-Coddington model).  One of the horizontal channels
passes straight through the nodes, the other one is subject to the
rules specified for the model of Fig.~5a.  The topological coupling of
the $\sigma$ model is then found to have the critical value
$\sigma_{xy}^{(0)} = 3 \times 1/2 \ ({\rm mod} \ 1) = 1/2$.  Thus the
last, doubly modified network model is critical and, on symmetry
grounds, lies in the quantum Hall universality class.  From what was
said above, it is also a suitable starting point for attempting to
deform to an integrable model.  \bigskip

\section{Summary}

Several messages result from the present paper.  First of all, a close
relation between two standard models of the integer quantum Hall
transition, namely the Chalker-Coddington model at its symmetric point
$p = 1/2$, and the supersymmetric formulation of Pruisken's nonlinear
$\sigma$ model at $\theta = 2\pi\sigma_{xy}^{(0)} = \pi \ ({\rm mod}\ 
2\pi)$, was established.  Let us put this result in the proper
context.  To be sure, it has been clear for a number of years now that
some sort of relation between these models ought to exist.  We know
that both are critical and belong to the same universality class, so
they cannot but describe the same physics at long wave lengths.
However, prior to our work, the understanding of the precise
connection between Chalker-Coddington and Pruisken was rather
indirect, relying on a double use of the anisotropic (or Hamiltonian)
limit.  The connection went as follows.  At one end, by taking the
Chalker-Coddington model and going to its anisotropic limit, D.H.~Lee
arrived at a network model consisting of an array of chiral modes with
alternating direction of motion.  At the other end, the anisotropic
limit of Pruisken's model was investigated, by an elaboration of the
work of Shankar and Read \cite{sr} on the ${\rm O}(3)$ nonlinear
$\sigma$ model.  It was argued in \cite{mrz_iqhe} that the Hamiltonian
limit of Pruisken's model at $\theta = \pi$ and strong coupling (small
$\sigma_{xx}^{(0)}$) is an antiferromagnetic superspin chain.  Now,
the array of chiral modes and the superspin chain are easy to relate
by conventional techniques.  At the level of the replica trick this
was done in \cite{DHLee}, the correct supersymmetric extension follows
from \cite{spinchain}.  In more detail, the functional integral
representation of the array of chiral modes maps on a nonlinear
$\sigma$ model with an alternating sum of Wess-Zumino terms.  (As we
have seen, this mapping is based on a $0+1$ dimensional analog of
nonabelian bosonization in $1+1$ dimensions.)  The latter, in turn,
coincides with the coherent-state path integral of the
antiferromagnetic superspin chain.

One disadvantage of the above way of relating the models is that one
does not have good control over the numerical value of the coupling
constant $\sigma_{xx}^{(0)}$.  In the present work, this uncertainty
was resolved by dealing directly with the {\it isotropic} models,
using a novel scheme devised in \cite{mrz_circular}.  What we have
shown is this.  Starting from the Chalker-Coddington model at $p =
1/2$ and {\it doing no more than an exact transformation followed by a
  continuum limit}, we arrive at Pruisken's model with $\theta = \pi$
and $\sigma_{xx}^{(0)} = 1/2$.  Or, in different words, the former
model is equivalent to a specific lattice discretization of the
latter.

Previously, Pruisken's model was thought to be associated primarily
with the white noise limit $l_c \ll l_B$, which is where Pruisken's
derivation starts from.  Recall, though, that in the course of
deriving his model, Pruisken made a saddle-point approximation to
eliminate the ``massive'' modes.  Naive use of this approximation
scheme is justified only in the limit of large $\sigma_{xx}^{(0)}$
(high Landau level).  In contrast, the present work makes {\it no}
such approximation.  The only assumption we needed was the dominance
of slowly varying fields in the functional integral, allowing us to
pass from the lattice to the continuum.  Thus, contrary to what might
have been expected, Pruisken's model at small $\sigma_{xx}^{(0)}$ is
actually associated more closely with the {\it high-field limit}, $l_c
\gg l_B$, as it is this limit that provides the microscopic
justification of the network model.  Note, however, that the ratio of
microscopic length scales $l_c / l_B$ is expected to be an irrelevant
parameter at a critical point with infinite correlation length.  Thus,
our result is not in conflict with Pruisken's derivation of the
$\sigma$ model model as a critical theory.

In my opinion, neither the Chalker-Coddington model nor Pruisken's
model hold much promise for an exact analytical solution in the near
future.  If so, the mapping of one model on the other is not yet a big
step forward.  The good news is that there are several useful
spinoffs.  D.K.K. Lee and Chalker suggested to model the random flux
problem (i.e., the motion of a single electron in a random magnetic
field) by a network with two channels per link and local ${\rm U}(2)$
gauge invariance.  Our mapping onto a nonlinear $\sigma$ model easily
extends to include this case.  The coupling constants of the continuum
field theory were found to be $\sigma_{xx}^{(0)} = \sigma_{xy}^{(0)} =
2 \times 1/2 \ ({\rm mod}\ 1) = 0$ at the symmetric point of the
random flux problem.  This is a massive theory with exponential decay
of all correlation functions are large scales.  Thus, contrary to
claims made in the literature, there is no room for truly extended
states in the random flux problem, at least not by slight deformation
away from the two-channel network model.  This conclusion had already
been reached in \cite{lee_chalker} and \cite{DHLee}.

Another spinoff helps us along in our quest to understand the integer
quantum Hall transition.  We observed that the SUSY reformulated
Chalker-Coddington network model has the structure of what is known as
a vertex model in statistical physics.  Motivated by information from
the theory of integrable systems, we then modified the network model
in several ways.  First, we abandoned the alternating direction of the
electron's motion along the horizontal and vertical straight lines of
the network.  Instead, we took the direction of motion to be constant
along every such line.  The mapping onto Pruisken's model indicates
that this modification changes the physics: the value of the
topological coupling now is $\sigma_{xy}^{(0)} = 0$, which corresponds
to a noncritical state.  We argued that criticality can be restored by
doubling the number of channels on all horizontal links (or on all
vertical links).  Alternatively, we can spatially separate the two
channels on horizontal links and return to a model with only one
channel per link, at the expense of doubling the size of the unit cell
in the vertical direction.

The resulting modified network model is critical, but no longer has a
justification from a microscopic picture of guiding center drift along
equipotentials.  Its virtue is that the corresponding supersymmetric
model is of a type for which systematic ways of constructing solutions
of the quantum Yang-Baxter equation are in principle available.
Whether our model can actually be deformed into one that is a Yang-Baxter
integrable, is a question whose answer lies far beyond the scope of
the present paper.

{\sl Acknowledgment.} Most of this work was done while the author was
visiting the Institute for Theoretical Physics, UCSB, Santa Barbara,
and the Aspen Center for Physics.  Useful discussions with John
Chalker, Matthew Fisher, Andreas Ludwig, Boris Muzykantskii and 
Nick Read are acknowledged.


\begin{thebibliography}{99}

\bibitem{huckestein} B. Huckestein, ``Scaling theory of the integer
  quantum Hall effect'', Rev. Mod. Phys. 67, 357-396 (1995).

\bibitem{trugman} S.A.~Trugman, ``Localization, percolation, and the
  quantum Hall effect'', Phys. Rev. B {\bf 27}, 7539-7546 (1983).

\bibitem{cc} J.T.~Chalker and P.D.~Coddington, ``Percolation, quantum
  tunneling, and the integer quantum Hall effect'', J. Phys. C {\bf
    21}, 2665-2679 (1988).

\bibitem{DHLee} D.H.~Lee, ``Network models of quantum percolation and
  their field-theory representation'', Phys. Rev. B {\bf 50},
  10788-10791 (1994).

\bibitem{ho_chalker} C.-M. Ho and J.T. Chalker, ``Models for the
  integer quantum Hall effect: the network model, the Dirac equation,
  and a tight-binding Hamiltonian'', cond-mat/9605073.

\bibitem{wegner} F.~Wegner, ``The mobility edge problem: continuous
  symmetry and a conjecture'', Z. Phys. B {\bf 35}, 207-210 (1979).

\bibitem{efetov} K.B. Efetov, ``Supersymmetry and theory of disordered
metals'', Adv. Phys. {\bf 32}, 53-127 (1983).  

\bibitem{pruisken} A.M.M. Pruisken, ``On localization in the theory of
  the quantized Hall effect: a two-dimensional realization of the
  $\theta$-vacuum'', Nucl. Phys. B {\bf 235}, 277-298 (1984).

\bibitem{pruisken1} A.M.M. Pruisken, ``Quasi particles in the theory
  of the integral quantum Hall effect (II).  Renormalization of the
  Hall conductance or instanton angle theta'', Nucl. Phys. B {\bf
    290}, 61-86 (1987).

\bibitem{llp} H. Levine, S.B. Libby and A.M.M. Pruisken, ``Theory of
  the quantized Hall effect (I)'', Nucl. Phys. B {\bf 240}, 30-40
  (1984).

\bibitem{haw} H.A.~Weidenm\"uller, ``Single electron in a random
  potential and a strong magnetic field'', Nucl.~Phys.~B {\bf 290},
  87-110 (1987).

\bibitem{Khmelnitskii} D.E.~Khmel'nitskii, ``Quantization of Hall
  conductivity'', JETP Lett.~{\bf 38}, 552-556 (1983).

\bibitem{mrz_iqhe} M.R. Zirnbauer, ``Towards a theory of the integer
  quantum Hall transition: from the nonlinear sigma model to superspin
  chains'', Ann. d. Physik {\bf 3}, 513-577 (1994); cond-mat/9410040.

\bibitem{perelomov} A.M. Perelomov, {\it Generalized coherent states
    and their applications} (Springer-Verlag, Berlin, 1986).

\bibitem{mrz_circular} M.R. Zirnbauer, ``Supersymmetry for systems
  with unitary disorder: circular ensembles'', J. Phys. A {\bf 29},
  7113-7136 (1996).

\bibitem{chari_pressley} V. Chari and A. Pressley, {\it A Guide to
    Quantum Groups} (Cambridge University Press, Cambridge, 1994).

\bibitem{vz} J.J.M.~Verbaarschot and M.R.~Zirnbauer, ``Critique of
    the replica trick'', J.~Phys.~A {\bf 17}, 1093-1109 (1985).

\bibitem{haw_mrz} H.A.~Weidenm\"uller and M.R.~Zirnbauer, ``Instanton
  approximation to the graded nonlinear sigma model for the integer
  quantum Hall effect'', Nucl.~Phys.~B {\bf 305}, 339-366 (1988).

\bibitem{rothstein} M.J. Rothstein, ``Integration on noncompact
  supermanifolds'', Trans. Am. Math. Soc. {\bf 299}, 387-396 (1987).

\bibitem{mrz_suprev} M.R. Zirnbauer, ``Riemannian symmetric
  superspaces and their origin in random matrix theory'', J. Math.
  Phys. {\bf 37}, 4986-5018 (1996). 

\bibitem{xrs} S. Xiong, N. Read, and A.D. Stone, ``Mesoscopic
  conductance and its fluctuations at non-zero Hall angle'', Phys.
  Rev. B {\bf 56}, 3982-4012 (1997).

\bibitem{lee_chalker} D.K.K.~Lee and J.T.~Chalker, ``Unified model for
  two localization problems: electron states in spin-degenerate Landau
  levels and in a random magnetic field'', Phys.~Rev.~Lett.~{\bf 72},
  1510-1513 (1994).

\bibitem{spinchain} L. Balents, M.P.A. Fisher, and M.R. Zirnbauer,
  ``Chiral metal as a ferromagnetic super spin chain'', Nucl. Phys. B
  {\bf 483}, 601-636 (1997).

\bibitem{MIT} C. Mudry, C. Chamon, and X.G. Wen, ``Two-dimensional
  conformal field theory for disordered systems at criticality'',
  Nucl. Phys. B {\bf 466}, 383-443 (1996).

\bibitem{grs} I.A. Gruzberg, N. Read, and S. Sachdev, ``Scaling and
  crossover functions for the conductance in the directed network
  model of edge states'', Phys. Rev. B {\bf 55}, 10593-10601 (1997).

\bibitem{fradkin} E. Fradkin, {\it Field theories of condensed matter
    systems} (Addison-Wesley, Redwood City, 1991).

\bibitem{baxter} R.J.~Baxter, {\it Exactly Solved Models in
    Statistical Mechanics} (Academic Press, London, 1982).

\bibitem{drinfeld} V.G. Drinfel'd, ``Quantum groups'', in Proceedings
  of the I.C.M., Berkeley 1986, Am. Math. Soc.  798-820 (1987).

\bibitem{sr} R.~Shankar and N.~Read, ``The $\theta = \pi$ nonlinear
  $\sigma$ model is massless'', Nucl.~Phys.~B {\bf 336}, 457-474
  (1990).
\end{thebibliography}
\end{document}